\documentclass{aastex63}

\usepackage{bm}
\usepackage{amsmath}
\usepackage{color}
\usepackage{graphicx}
\usepackage{ulem}
\defcitealias{2023ApJ...950...72K}{Paper~I}
\defcitealias{2023MNRAS.526.2717W}{Paper~II}
\received{June 1, 2019}
\revised{January 10, 2019}
\accepted{\today}
\submitjournal{ApJ}

\shorttitle{Dynamical structure in AGN sub-pc dusty outflows}
\shortauthors{Kudoh et al.}
\begin{document}

%%%%%%%%%%%%%%%%%%%%%%%%%%%%%%%%%%%%%%%%%%%%%%%%%%%%%%%%
\title{Multiphase Gas Nature in the Sub-parsec Region of the Active Galactic Nuclei. III. Eddington Ratio Dependence on the Structures of Dusty and Dust-free Outflows}
%%%%%%%%%%%%%%%%%%%%%%%%%%%%%%%%%%%%%%%%%%%%%%%%%%%%%%%
\correspondingauthor{Kudoh Yuki}
\email{yuki.kudoh@astrophysics.jp}

%%%%%%%%%%%%%%%%%%%%%%%%%%%%%%%%%%%%%%%%%%%%%%%%%%%%%%%%
%\author[0000-0002-0786-7307]{Greg J. Schwarz}
\author[0000-0003-0548-1766]{Kudoh Yuki}
\affiliation{
Astronomical Institute, Tohoku University,\\
6-3 Sendai, Miyagi 980-8578, Japan}

\author[0000-0002-8779-8486]{Wada Keiichi}
\affiliation{
Graduate School of Science and Engineering, Kagoshima University,\\
1-21-35 Korimoto, Kagoshima 890-0065, Japan}

\author[0000-0003-2535-5513]{Kawakatu Nozomu}
\affiliation{
Faculty of Natural Sciences, National Institute of Technology, Kure College, \\
2-2-11 Agaminami, Kure, Hiroshima 737-8506, Japan}

\author[0000-0002-6236-5270]{Nomura Mariko}
\affiliation{
Graduate School of Science and Technology, Hirosaki University, \\
3 Hirosaki, Aomori 036-8561, Japan}


%%%%%%%%%%%%%%%%%%%%%%%%%%%%%%%%%%%%%%%%%%%%%%%%%
\begin{abstract}

We investigated the influence of the Eddington ratio on
sub-parsec-scale outflows in active galactic nuclei (AGNs) with supermassive black holes (SMBHs) masses of $10^7 M_{\odot}$ using two-dimensional radiation
hydrodynamics simulations. 
When the range of Eddington ratio, $\gamma_{\rm Edd} > 10^{-3}$ 
, the radiation force exceeds the gas pressure, leading to
stronger outflows and larger dust sublimation radius.
Although the sub-parsec-scale outflows is a time-dependence phenomena, our simulations demonstrated that the radial distributions can be well explained by the steady solutions of the spherically symmetric stellar winds.
The dynamic structure of sub-parsec-scale outflows is influenced by the dust sublimation radius and the critical radii determined by the dynamical equilibrium condition. 
Although significantly affecting the outflow velocity, the Eddington ratio exerts minimal effects on temperature and number density distribution. 
Furthermore, our analytical solutions highlight the importance of the dust sublimation scale as a crucial determinant of terminal velocity and column density in dusty outflows. 
Through comparisons of our numerical model with the obscuring fraction observed in nearby AGNs, we revealed insights into the Eddington ratio dependence and the tendency towards the large obscuring fraction of the dusty and dust-free gases. 
The analytical solutions are expected to facilitate an understanding of the dynamical structure and radiation structures along the line of sight and their viewing angles from observations of ionized outflows.

\end{abstract}



\section{Introduction} \label{sec:intro}
The unified picture of active galactic nuclei (AGN) provides a framework for understanding the obscuration of gas and dust interacting with AGN radiation, represented as the column density of a dust torus located outside a broad-line region (BLR) surrounding a supermassive black hole (SMBH) \citep[][]{1993ARA&A..31..473A, 1995PASP..107..803U}. 
The AGN luminosity is determined by the gas supply onto a SMBH. 
However, as the mass accretion rate may be affected by AGN outflows, it is essential to investigate the physical process of AGN outflows to evaluate the realistic growth rate of SMBHs \citep[e.g.,][]{1998A&A...331L...1S,2012MNRAS.425..605F,2015ARA&A..53..115K,2020MNRAS.494.3616N}.

Recent multi-wavelength observations highlight that AGN outflows represent complex environments with multiphase gases, e.g. ultra-fast outflows (UFOs), warm absorbers (WAs), ionized, atomic, and molecular outflows (see, \citealt{2018NatAs...2..198H,2021agnf.book.....C}).
In the nearby Seyfert Circinus galaxy, \citet{2023Sci...382..554I} revealed the multiphase nature of outflows and inflows from ten-parsec to sub-parsec scales using the Atacama Large Millimeter/sub-millimeter Array (ALMA).  
In addition, the infrared interferometry observations have clarified the dust thermal radiation elongating to the polar region \citep{2012ApJ...755..149H,2022ApJ...940...28K,2022ApJ...940L..31L}. 
Observational studies have investigated that the parameter determining AGN activity is the Eddington ratio $\gamma_{\textrm{Edd}}$, which is the bolometric luminosity $L_{\textrm{bol}}$ divided by the Eddington luminosity $L_{\textrm{Edd}}$ proportional to the SMBH mass $M_{\textrm{SMBH}}$,
\begin{equation}
\gamma_{\textrm{Edd}}= \frac{ L_{\textrm{bol}} }{ L_{\textrm{Edd}} }  \propto  \frac{  L_{\textrm{bol}} }{ M_{\textrm{SMBH}} }.
\label{eq:Eddingtonratio}
\end{equation}

In contrast to the cold dust in the torus, the dust in the polar region indicates high temperatures $\gtrsim 1000$ K \citep{2013ApJ...771...87H,2014A&A...563A..82T}.
The Infrared (IR) luminosity from this thermal emission correlates with the X-ray luminosity  \citep{2017MNRAS.469..110G,2022A&A...667A.140G}.
Additionally, there have been discussions based on the dependence of Eddington ratio and hydrogen column density \citep{2019MNRAS.489.2177A,2021A&A...652A..99A,2023ApJS..265...37Y}.
The polar dust is spatially associated with ionized outflow, forming horn-like structures \citep{2016ApJ...822..109A,2023MNRAS.519.3237S}. 
The outflowing gas exhibits peak velocities $>100$ km s$^{-1}$ and number densities of approximately $10^3$ cm$^{-3}$, which are related with the AGN luminosity \citep{2017A&A...601A.143F,2020MNRAS.498.4150D,2023A&A...679A..84M}.
Understanding the central region of the horn, that is the sub-parsec scale, is crucial for the origin of formation mechanism.

The gas and dust in observed AGNs is characterized by obscuring fraction $f_{\rm obs}$
 \footnote
{
The fraction of gas covering or obscuration is often expressed as covering fraction, covering factor, or similar terms. 
These generally relate to partial absorption along the line of sight or the absorption/emission of gas covering a certain spherical surface. 
The factors introduced in AGN observations to adjust luminosity involve uncertainties related to the gas distribution \citep[see in detail, ][]{2015ARA&A..53..365N}.
} 
.
\cite{2013MNRAS.435.1840R} clarified the X-ray obscuring fraction, $f_{\mathrm{obs, X}}$, by examining the anti-correlation between the Fe K$\alpha$ equivalent width with luminosity, known as the X-ray Baldwin effect \citep{1993ApJ...413L..15I}. 
$f_{\mathrm{obs, X}}$ represents the obscured region of the torus according to the unified model. 
On the other hand, the assumption that the dusty torus re-emits a portion of the AGN luminosity in the IR allows for the conversion of the dust obscuring fraction, $f_{\mathrm{obs, IR}}$ \citep{2007A&A...468..979M,2008ApJ...679..140T}.

Recent studies have reported a higher $f_{\rm obs, X}$ \citep[e.g., ][]{2014ApJ...786..104U,2015MNRAS.451.1892A,2015ApJ...802...89B} compared to $f_{\rm obs, IR}$ \citep[e.g.,][]{2007A&A...468..979M,2008ApJ...679..140T,2013ApJ...777...86L,2016ApJ...819..123N,2018ApJ...862..118Z}, which is discussed by \citet{2020ApJ...897....2T,2021ApJ...906...84O,2021A&A...651A..91E}.
$f_{\rm obs, IR}$ captures only dusty gas, whereas $f_{\rm obs, X}$ contains additional dust-free gas.
The difference between $f_{\rm obs, IR}$ and $f_{\rm obs, X}$ suggests that the inner part of the dusty torus is more widely obscured by dust-free gas, which is destroyed by dust sublimation \citep{2015ApJ...806..127D,2016A&A...586A..28B,2019ApJ...870...31I,2022MNRAS.516.2876M,2024MNRAS.532..666M}. 
These are dependent on the Eddington ratio, with a tendency for smaller obscuring fraction with high Eddington ratio \citep{2021ApJ...912...91T,2022ApJ...939L..13A}.
\citet{2017Natur.549..488R,2023ApJ...959...27R,2024MNRAS.529.3610V} discussed that this trend implied gas being blown out by radiative feedback.

In this paper, we define the obscuring fraction $f_{\mathrm{obs}}$ based on geometry. 
This is the fraction of a hemisphere ($2\pi$) covered by an angle $\theta$ from the disk mid-plane.
The interpretation of observed $f_{\mathrm{obs}}$ in relation to gas distribution is discussed in \citep{2012ApJ...747L..33E,2008ApJ...685..160N,2016MNRAS.458.2288S}.

The importance of radiation-driven dusty outflow is demonstrated by hydrodynamic simulations \citep[e.g.,][]{2012ApJ...759...36R,2012ApJ...758...66W,2015ApJ...812...82W,2016MNRAS.460..980N,2017ApJ...843...58C,2021ApJ...920...30N,2023MNRAS.525.2668S}.
Notably, \citet{2012ApJ...758...66W, 2015ApJ...812...82W} revealed a fountain mechanism wherein the multiphase gas blew up from the disk within a few parsecs, subsequently falling back to the disk plane at a scale where the cooling became effective.
Their dynamic model at the parsec scale successfully explained the spectral lines of the outflowing atomic/molecular gases \citep{2018ApJ...867...48I,2023Sci...382..554I} and the polar dust emission \citep{2014MNRAS.445.3878S, 2016ApJ...828L..19W}.
Considering steady winds from scales smaller than the their spatial resolution ($\sim 0.1$ pc), \citet{2019ApJ...876..137W,2020ApJ...897...26W} showed that the dynamical structure of dusty outflows varied with the Eddington ratio and radiation anisotropy.
To explore the origin of the dusty outflow at $\gamma_{\rm Edd}=0.1$, by spatially resolving the dust sublimation scale, \citet[][]{2023ApJ...950...72K} (Paper I) investigated the time variability in the shape of outflowing shells and dust sublimation radii.
The timescale of variability within a few decades can impact X-ray polarization \citep{2023ApJ...958..150T} and ionized emission lines \citep[][ Paper~II]{2023MNRAS.526.2717W}.
Moreover, the dust sublimation region was filled with dense, dust-free gas owing to the destruction of dusty gas flowing out from the disk.

The sub-parsec-scale outflow plays a crucial role in understanding the dynamics of dust-free and dusty gases, contributing to the obscured structure covering the central nuclei.
In this paper, we studied the dependence of the Eddington ratio on sub-pc-scale radiation feedback and its wind structure through the numerical simulations based on \citetalias{2023ApJ...950...72K}. 
We aimed to establish an analytical model of the Eddington ratio based on the simulations results.
We answer the following questions, as explained by the radiative feedback and its dynamics:
(1) How does the radiation feedback to the sub-parsec-scale gas depend on the Eddington ratio?;
(2) What dynamics govern the radial distribution of the dynamical outflow?;
(3) What is the value of Eddington ratio at which radiation feedback is activated on the sub-parsec scale?;
(4) How does the dynamical dusty outflow explain the obscuring fraction of IR and X-ray observations?

This paper is organized as follows. 
In \S \ref{sec:model}, we describe the solving equations and physical models common to our simulations in \S\S \ref{sec:sub21} and the models of AGN luminous sources parameterized by the Eddington ratio in \S\S \ref{sec:sub22}.
We present the numerical results of the radiation-driven outflow and the time-averaged radial profile according to the Eddington ratio in \S \ref{sec:results}. 
We establish a dynamic model that reproduces the time-averaged outflow velocity by analytical solutions in \S \ref{sec:analytical}.
\S\S \ref{sec:sub51} discusses the application of observed outflow velocity using the analytical solutions. 
We discuss the column density for dusty and dust-free gases in \S\S \ref{sec:sub52}, and its $f_{\rm obs}$ compared to the IR and X-ray observations in \S\S \ref{sec:sub53}.   
In addition, \S\S \ref{sec:sub54} and \S\S \ref{sec:sub55} rough out the ionization parameter related to X-ray winds and driving forces classification. 
Finally, \S \ref{sec:summary} summarizes our findings related to the dynamical dusty outflow dependence on the Eddington ratio.

%%%%%%%%%%%%%%%%%%%%%%%%%%
\begin{figure}[h!]
%\begin{interactive}{animation}{fig4.mp4}
\epsscale{0.8}
\plotone{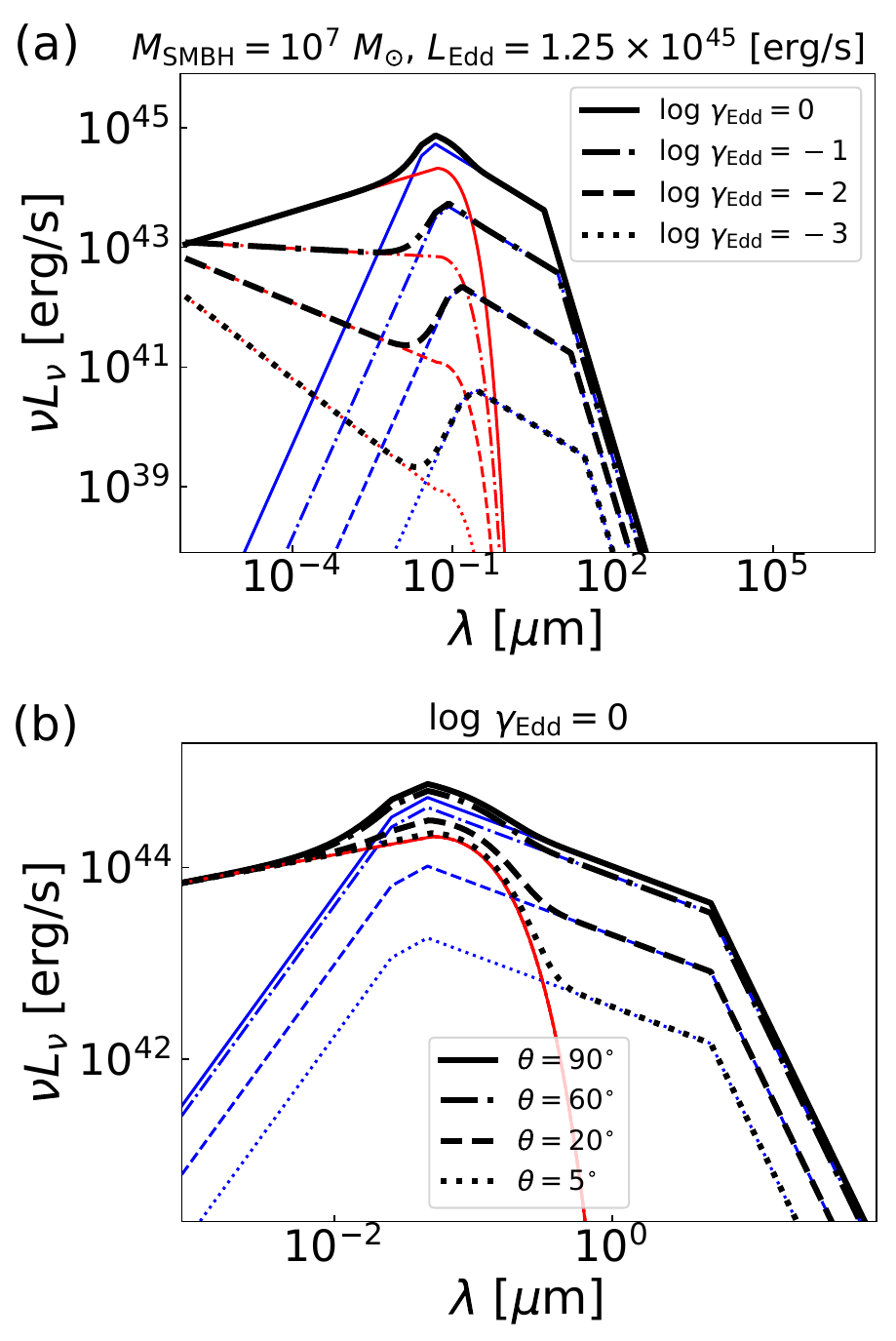}
%\end{interactive}
\caption{
(a) Models of the input AGN SEDs. The black lines are total SEDs
for Eddington ratio ($\gamma_{\rm Edd}$). The blue and red lines are 
the accretion disk and the disk corona components, respectively. 
(b) Same as (a), but for the angular dependence. The black lines are $\theta=90^{\circ}$ (pole-on), $60^{\circ}, 20^{\circ}$, and $5^{\circ}$ (mostly edge-on). 
\label{fig:1}}
\end{figure}
%%%%%%%%%%%%%%%%%%%%%%%%%%

\section{Model} \label{sec:model}
We investigated the dependence of the Eddington ratio through numerical simulations of sub-parsec-scale radiation-driven dusty outflow induced by radiation sources. 
We modeled an accreting dusty disk that was irradiated from an anisotropic central radiation field with a central black hole mass ($10^7 M_{\odot}$) in axisymmetric cylinder coordinates $(R,z)$. 
The simulations in this study were performed using the public code \verb+CANS++ \citep{2019PASJ...71...83M}, which includes ray-trace modules for radiative heating and radiation force from frequency-dependent radiation sources. 
To simplify the numerical setup and the physical model for dust and gas, we adopted the same setup as that mentioned in \citetalias{2023ApJ...950...72K}. 
We provide a brief description of the common model and then introduce the updated SED model incorporating the Eddington ratio.

\subsection{Basic equations and numerical setup} \label{sec:sub21}
The equations that were solved included gas pressure, radiation force, and gravity, which are expressed as follows:
%continuum
\begin{eqnarray}
\frac{\partial \rho}{\partial t} + \bm{\nabla} \cdot \left[\rho \bm{v} \right]  =0,  
\label{eq:mass}
\end{eqnarray}
% equation of motion
\begin{eqnarray}
 \displaystyle \frac{\partial \rho \bm{v}}{\partial t}
 + \bm{\nabla} \cdot \left[ \rho  \mathbf{vv} + {P}_{\rm g} \mathbf{I}  \right]  
 =  \bm{f}_{\rm rad} + \bm{f}_{\rm grav} + \bm{f}_{\rm vis},
\label{eq:momentum}
\end{eqnarray}
% energy equation
\begin{eqnarray}
\displaystyle \frac{\partial e}{\partial t}
+ \bm{\nabla} \cdot \left[ \left( e+ P_{\rm g}  \right) \bm{v}  \right]
 = - \rho {\cal L} + \bm{ v} \cdot \bm{f}_{\rm rad} + \bm{ v} \cdot \bm{f}_{\rm grav} + W_{\rm vis},
 \label{eq:energy}
\end{eqnarray}
where $\rho$ is the total density of gas and dust, assuming the dust-to-gas mass ratio $\delta_{\rm dg} =0.01$, $\bm{v}$ is the velocity, and $P_{\rm g}$ is the gas pressure.
The total energy density $e$ is expressed as $P_{\rm g}/\left( \gamma -1 \right)+ \rho |\bm{v}|^2/2$ with the specific heat ratio $\gamma=5/3$.
 $\bm{f}_{\rm grav} =  - \bm{e}_r \rho G M_{\rm SMBH}/r^2 $ at radius $r=\sqrt{R^2+z^2}$ is the gravitational force of the SMBH mass with $M_{\rm SMBH}= 10^7 M_{\odot}$ with the direction away from the center $\bm{e}_{r}$.
$\bm{f}_{\rm vis}$ and $W_{\rm vis}$ are the viscous force and the viscous heating \citep{2005ApJ...628..368O} with the assumption of the $\alpha$-viscous model as $\alpha=0.1$ \citep{1973A&A....24..337S}.  
These equations describe an axisymmetric system in the $R$-$z$ plane, while also considering angular momentum conservation.
${\cal L}$ denotes the net heating/cooling rate per unit mass \citep{2009ApJ...702...63W}.
$\bm{f}_{\rm rad} = \int\nabla \cdot  \left( F_{\nu} \bm{e}_r \right) d\nu $ represents the radiation forces integrated over the frequency $\nu$ within $10^{16}$ -- $10^{22}$ Hz. 
The radiation field from the accretion disk contributes to both radiation force and radiative heating, which are treated as emanating from a point source.
As reported by \cite{2016MNRAS.460..980N}, the radiation from the dusty gas disk on sub-parsec to parsec scales is inefficient. 
Therefore, we neglect the radiation field from the dusty gas on these scales.  
$F_{\nu}$ is the central radiation flux per unit frequency including the effect of extinction by optical depth $\tau$ determined by dust opacity $\kappa_{\rm d}$ and Thomson scattering $\kappa_{\rm T}$.
When dust is destroyed, $\kappa_{\rm d}$ becomes zero.
The conditions are required are as follows: 
(1) the sputtering timescale \citep{1995ApJ...448...84T} is shorter than the dynamical timescale, and 
(2) the dust temperature exceeds the dust sublimation temperature $T_{\rm sub} =1500$ K. 
We calculated the dust temperature assuming a local thermal equilibrium with the incoming radiation flux $F_{\nu}$. 
Further details regarding the central radiation flux $F_{\nu}$ are provided in the next subsection.

The computational grids and domain in the cylindrical coordinate were $(N_R, N_z) = (210, 1286)$ and $10^{-4} < R < 2$ pc, $-2 < z < 2$ pc.
The grid structures were uniform with $\Delta R = \Delta z = 5 \times 10^{-4}$ pc for $R < 3.8 \times 10^{-2}$ pc and $|z| < 9.6 \times 10^{-2}$ pc, and non-uniform for $R > 3.8 \times 10^{-2}$ pc and $|z| > 9.6 \times 10^{-2}$ pc, stretched up to $\Delta R / R \lesssim 0.05$.
The boundary conditions were symmetrical for $\rho, P_{\text{g}}$, and $v_z$ and antisymmetrical for $v_R$ and $v_z$ with respect to the $z$-axis, while the remaining were set as outflow conditions.
In the central region where $r \leq 2 \times 10^{-3}$ pc, we adopted the absorbed boundary conditions with $\rho = 10^{-28}$ g cm$^{-3}$, $T_{\text{g}} = 10^4$ K, and $v_{\varphi} = 0$, respectively.
The initial density distribution in the Keplerian disk is expressed as
\begin{equation}
\rho(R,z) = \rho_0 \left( \frac{R}{R_0} \right)^{-p} \exp \left( \frac{|z|}{H(R)} \right)
\end{equation}
where the mid-plane density at $R_0 = 0.01$ pc is $\rho_0 = 10^{-10}$ g cm$^{-3}$ with a radial power $p = 3$ \citep[e.g.,][]{Kawaguchi2003}, and the disk scale height is $H(R) = 5 \times 10^{-3} R$.
Moreover, the disk temperature is constant, $T_{\textrm{gas}} = 100$ K.

\subsection{AGN radiation sources on the Eddington ratio} \label{sec:sub22}

Figure \ref{fig:1}a shows the four SED models from $10^{-6} \mu$m to $10^6 \mu$m for different Eddington ratio: $\log \gamma_{\textrm{Edd}} =0, -1, -2,$ and $-3$. 
The central radiation source of AGN was modeled by an accretion disk and its corona.
The blue curves of Figure \ref{fig:1}a showed the accretion disk SED $L_{\nu}^{\mathrm{AD}}$ \citep[see detail in][]{2005A&A...437..861S,2011MNRAS.415..741S}, as the Shakura-Sunyaev disk \citep{1973A&A....24..337S}.
Since this disk geometry is thin, radiation flax exhibits anisotropy dependence (e.g., \citealt{1987MNRAS.225...55N}),
$
f_{\mathrm{AD}}(\theta) = \sin \theta (1+ 2 \sin \theta),
$
where $\theta$ is defined in the range from 0 to $\pi/2$ and represents the inclination angle between the equatorial plane and the incoming radiation.
$f(\theta)$ has the factors for $\sin \theta$ of the change in the projected surface area and $(1+ 2 \sin \theta)$ of the limb darkening \citep[see also,][]{1985A&A...143..374S}.
We employed $F_{\nu} = L_{\nu} f_{\mathrm{AD}}(\theta)$ with angle dependence shown in Figure \ref{fig:1}b, while Figure \ref{fig:1}a is the case of $\theta =90^{\circ}$.
Figure \ref{fig:1}a also depict the red curves of disk corona SED following $L_{\lambda}^{\mathrm {corona}} \propto \lambda^{\Gamma-3}$ \citep[e.g.,][]{2016MNRAS.460..980N}.
The photon index $\Gamma$ is statistically correlated with the X-ray spectrum \citep{2013MNRAS.433.2485B}, expressed as $\Gamma=0.32 \log \gamma_{\mathrm{Edd}} +2.27$.
The disk corona was assumed to be spherically symmetric, rendering it independent of $\theta$, as shown in Figure \ref{fig:1}b.

To obtain the bolometric luminosity, i.e. $\gamma_{\mathrm{Edd}} L_{\mathrm{Edd}} = L_{\nu}^{\mathrm{AD}}$ + $L_{\nu}^{\mathrm{corona}}$, the relative magnitude of the disk and its corona must be determined.
We adopted the ratio $L_{\nu}^{\mathrm{AD}} / L_{\nu}^{\mathrm{corona}}$ as the flux ratio of 2500 \AA (ultraviolet) for  $L_{\nu}^{\mathrm{AD}}$ and 2 keV (X-ray) for $L_{\nu}^{\mathrm{corona}}$, $\alpha_{\mathrm{OX}}$, which is expressed as $\alpha_{\mathrm{OX}}=0.13 \log \gamma_{\mathrm{Edd}}+1.39$ \citep{2021ApJ...910..103L}.
Hence, the magnitude of these SEDs at $\theta=90^{\circ}$ are normalized as the bolometric luminosity.

\section{Numerical results} \label{sec:results}
%%%%%%%%%%%%%%%%%%%%
\begin{figure*}[h!]
%\begin{interactive}{animation}{fig4.mp4}
\epsscale{1.0}
\plotone{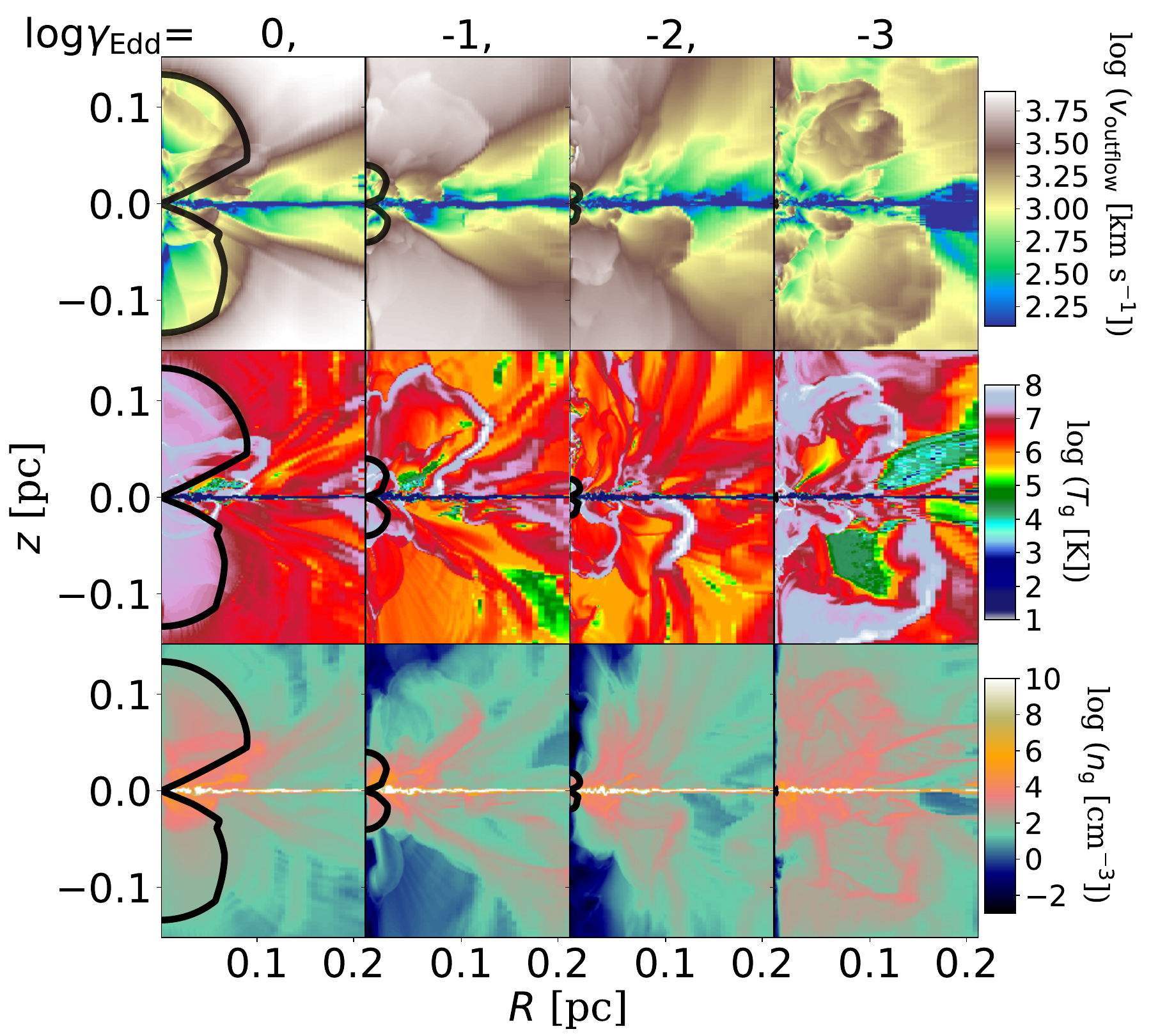}
%\end{interactive}
\caption{
From top to bottom, the spatial distributions denote the magnitude of outflow velocity, gas temperature, and number density for four models with different Eddington ratios, $\log \left(\gamma_{\rm Edd} \right)=0, -1, -2,$ and $-3$, respectively, at $t=4180$ years. 
The black curves represent the dust sublimation radius, which is invisible for $\log(\gamma_{\rm Edd}) = -3$.
\label{fig:2}
}
\end{figure*}
%%%%%%%%%%%%%%%%%%%% 

Figure \ref{fig:2} presents an overview of sub-parsec-scale outflows categorized by the Eddington ratio. 
These snapshots are captured at $t=4180$ years, thereby allowing sufficient time for the intermittent outflowing shells to demonstrate quasi-steady behavior. 
The black contours represent the dust sublimation radius at the dust temperature of $T_{\rm d} = 1500$ K. 
In the top panels of Figure \ref{fig:2}, the spatial distributions are shown as the outflow velocity defined as $v_{\rm outflow} = \sqrt{v_R^2+v_z^2}$. 
Outside the black contours, the region where the outflow velocity exceeds $10^4$ km s$^{-1}$ expands in response to the radiation strength corresponding to the Eddington ratio.
In addition, the outflow velocity is higher in the polar regions owing to radiation anisotropy (see also Figure \ref{fig:1}b). 
Conversely, within the dust sublimation radii, the outflow velocity is lower than the outer region, and in certain places, it dropped below the escape velocity of 
$
v_{\rm esc} \sim 2 \times 10^3 \text{ km s}^{-1} \left( M_{\rm SMBH}/ 10^7 M_{\odot} \right)^{1/2} \left(r/0.01 \text{ pc} \right)^{-1/2}
$.  
In this region, the high-temperature, dense gas exists at $\sim 10^8$ K and $\gtrsim 10^3$ cm$^{-3}$. 
However, apart from the size of this contours, the differences in the temperature and density owing to the Eddington ratio remain unclear.

The maximum size of the contours decreases from $\gamma_{\rm Edd}=1$ to $\gamma_{\rm Edd}=10^{-3}$, with values of 0.135, 0.041, 0.019, and 0.004 pc, scaling approximately with $\gamma_{\rm Edd}^{1/2}$
This scaling is consistent with the estimation of the local thermal equilibrium between the absorption of the radiation flux from the central sources and the dust thermal emission (see also Equations \ref{eq:r_sub_Edd}),
\begin{equation}
r_{\rm sub} = \sqrt{ \frac{ L_{\rm UV}}{4 \pi \sigma_{\rm SB}  T_d^{4} } } \propto M_{\rm SMBH}^{1/2} \left( \frac{L_{\rm UV}}{L_{\rm bol}}\right)^{1/2} \gamma_{\rm Edd}^{1/2}, 
\label{eq:r_sub}
\end{equation}
where $L_{\rm UV}$ is the UV luminosity and $\sigma_{\rm SB}$ is the Stefan-Boltzmann constant. 
The non-spherical shape is attributed to the radiation anisotropy and the shielding effect of the outflow itself. 
Owing to the time variability of the outflow, the dust sublimation radius is asymmetrical with respect to the equatorial plane and varies over time (see also Fig. 8 in \citetalias{2023ApJ...950...72K})

The flow pattern appeared as intermittent shells with shocks in the case of $\gamma_{\rm Edd}=0.1$.
\citetalias{2023ApJ...950...72K} reported that the spherical shell inside the dust sublimation radius was deformed by anisotropic radiation, resulting in an hourglass shape, as a result of inflow-induced outflow.
The sequential eruption of shells creates a pattern resembling the unfolding petals of a lotus flower.
Time variability is approximiately a few years \citep[see also,][]{2023MNRAS.526.2717W,2023ApJ...958..150T}.
These flows are launched from the surface of the dense dusty disk located within a few $10^{-3}$ pc driven by the radiative heating from irradiation.
In addition, they are accelerated under the influence of radial radiation forces, which are dependent on radiation anisotropy (blue lines in Figure \ref{fig:1}b).

%%%%%%%%%%%%%%%%%%%%
\begin{figure*}[h!]
%\begin{interactive}{animation}{fig4.mp4}
\epsscale{1.0}
\plotone{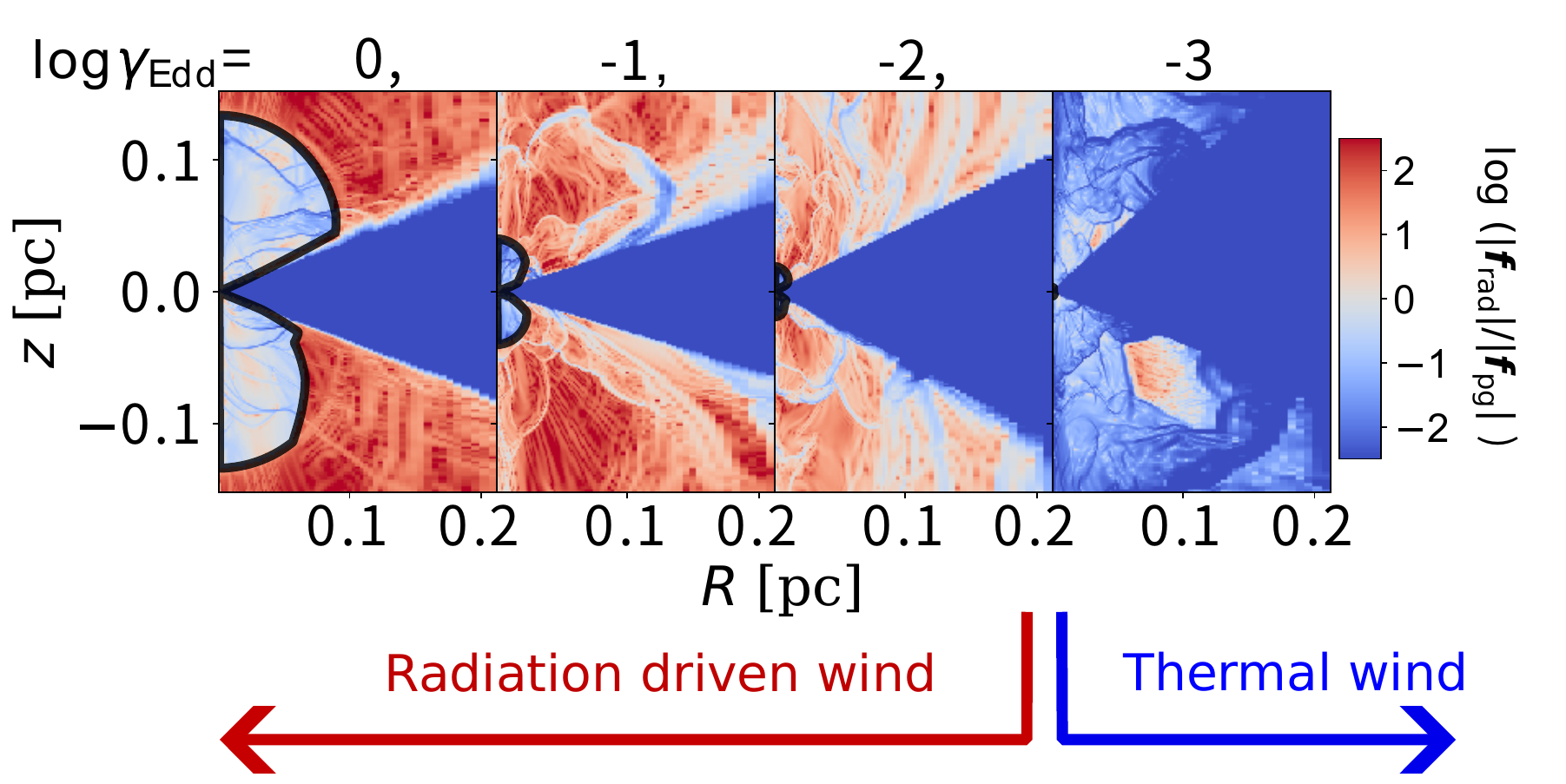}
%\end{interactive}
\caption{
Same snapshots as in Figure \ref{fig:2}, albeit showing the distribution of the ratio of the radiation force to gas pressure force. 
Red regions indicate where radiation force dominates, while blue regions indicate the dominance of gas pressure. 
In the model with $\log \left(\gamma_{\rm Edd} \right)=-3$, the thermal wind is primarily driven by gas pressure, contrasting with the radiation-driven wind observed in the other three models with larger $\gamma_{\rm Edd}$.
\label{fig:3}}
\end{figure*}
%%%%%%%%%%%%%%%%%%%%
To clarify the Eddington ratio dependence of outflow, Figure \ref{fig:3} shows the force ratio of radiation to gas pressure. 
The physical mechanisms driving the winds notably differ between the thermal wind, determined by gas pressure (blue region), for $\log \gamma_{\text{Edd}}=-3$, and the radiation-driven wind (red region) for $\log \gamma_{\text{Edd}} \geq -2$.
Figure \ref{fig:3} also shows the regions wherein the gas pressure exceeds the radiation force, independent of the Eddington ratio. 
One such region is located around $z=0$, attributed to the radiation absorption by the dense dusty gas disk. 
Moreover, the opacity of the dust-free gas within the dust sublimation radius is approximately 500 times smaller than that of dusty gas, resulting in the weakening of radiation force.
Even in regions dominated by radiation force at $\log \gamma_{\text{Edd}} > -2$, the gas pressure can increase within the dense shells. 
These shells are pushed outward by the radiation force and compressed by the shock, resulting in enhanced gas pressure (see also Figure \ref{fig:2}c).

%%%%%%%%%%%%%%%%%%%%
\begin{figure}[h!]
%\begin{interactive}{animation}{fig4.mp4}
\epsscale{1.0}
\plotone{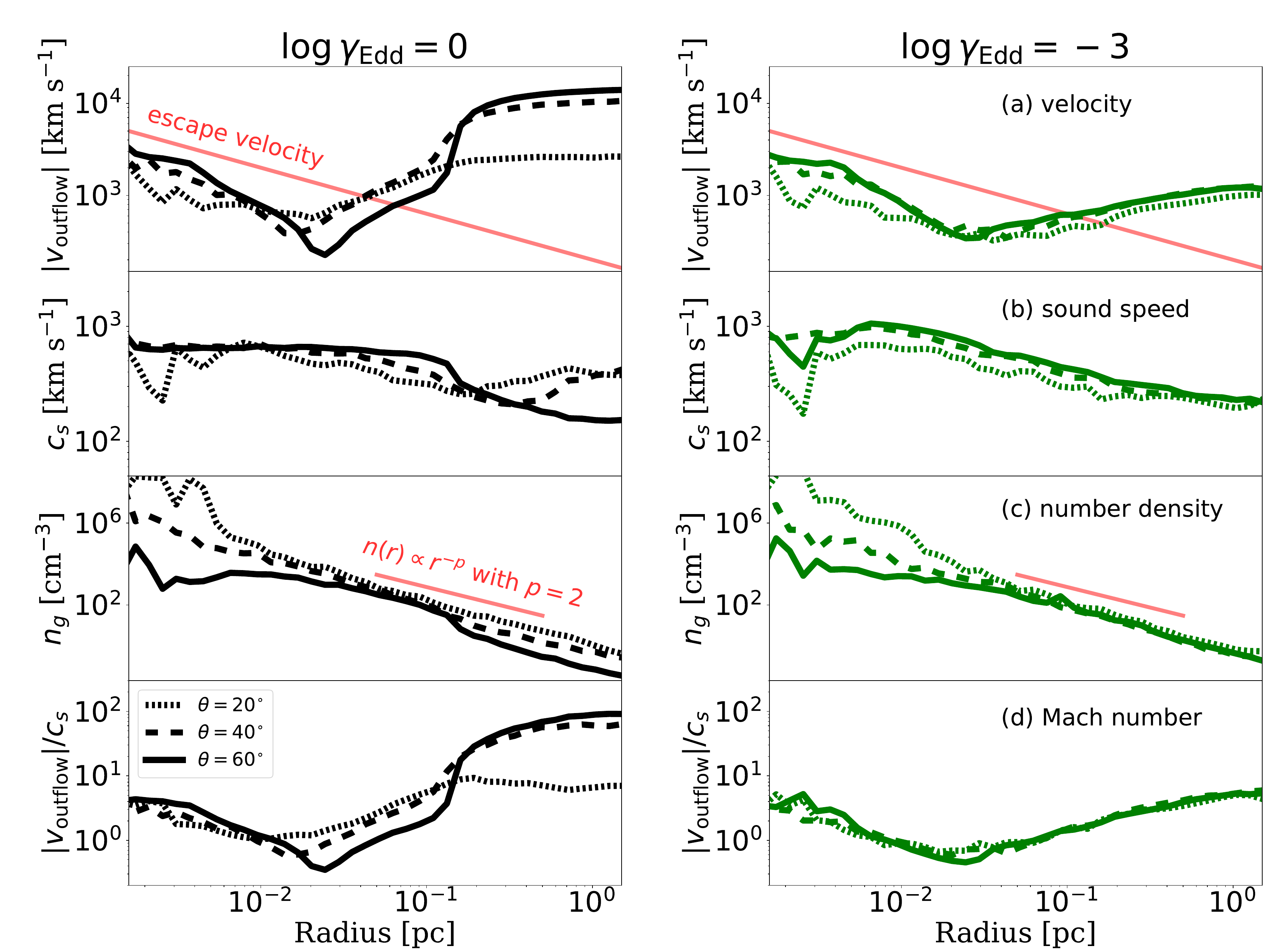}
%\end{interactive}
\caption{
Radial profiles comparing the radiation-driven wind at $\log (\gamma_{\rm Edd})=0$ (left) and the thermal wind at $\log (\gamma_{\rm Edd})=-3$ for various angles, $\theta=20$ (dotted line), $\theta=40$ (dashed line), and $\theta=60$ (solid line).
These profiles 
are obtained by
averaging snapshots over the period $4180 < t < 5573$ years, equivalent to 50 rotational times at $R=0.01$ pc. 
Panels (a) -- (d) show the outflow velocity, sound speed, number density, and Mach number, respectively.
Red solid lines in the panels (a) denote the escape velocity, while in the panels (c) indicate the scaling with $p=2$ of $n(r) \propto r^{-p}$.
\label{fig:4}}
\end{figure}
%%%%%%%%%%%%%%%%%%%%

Figure \ref{fig:4} shows the $\theta$--dependency of the outflow velocity, sound speed, number density, and Mach number for the radiation-driven wind at $\log \left( \gamma_{\rm Edd} \right)= 0$ and the thermal wind at $\log \left( \gamma_{\rm Edd} \right)= -3$. 
These quantities are plotted as time-averaged radial profiles, based on an average over 50 rotation periods at $r=0.01$ pc. 
These profiles represent at the averaged trajectory of the high-density shocked shells shown in Figure \ref{fig:2}.

Concerning the radiation-driven wind (left panels of Figure \ref{fig:4}), both the number density and temperature indicate similar to that of the thermal wind (right panels); however, there are discrepancies in the outflow velocity. 
In particular, the velocities at $r=1$ pc decrease with smaller $\theta$. 
In both cases, the number density can be fitted as a power law with $n(r) \propto r^{-p}$, where $p\sim 2$ for $r \gtrsim 0.1$ pc, while for $r \lesssim0.1$ pc, $p$ increases as the angle decreases from 60$^{\circ}$ to 20$^{\circ}$. 
We found that the power of the number density varies inside the point where the outflow velocity becomes below the escape velocity.
Upon examining the Mach number (panel d) for both wind types, it is evident that the profiles are primarily influenced by the outflow velocity, with a minor contribution from the sound speed.

%%%%%%%%%%%%%%%%%%%%
\begin{figure}[h!]
%\begin{interactive}{animation}{fig4.mp4}
\epsscale{1.0}
\plotone{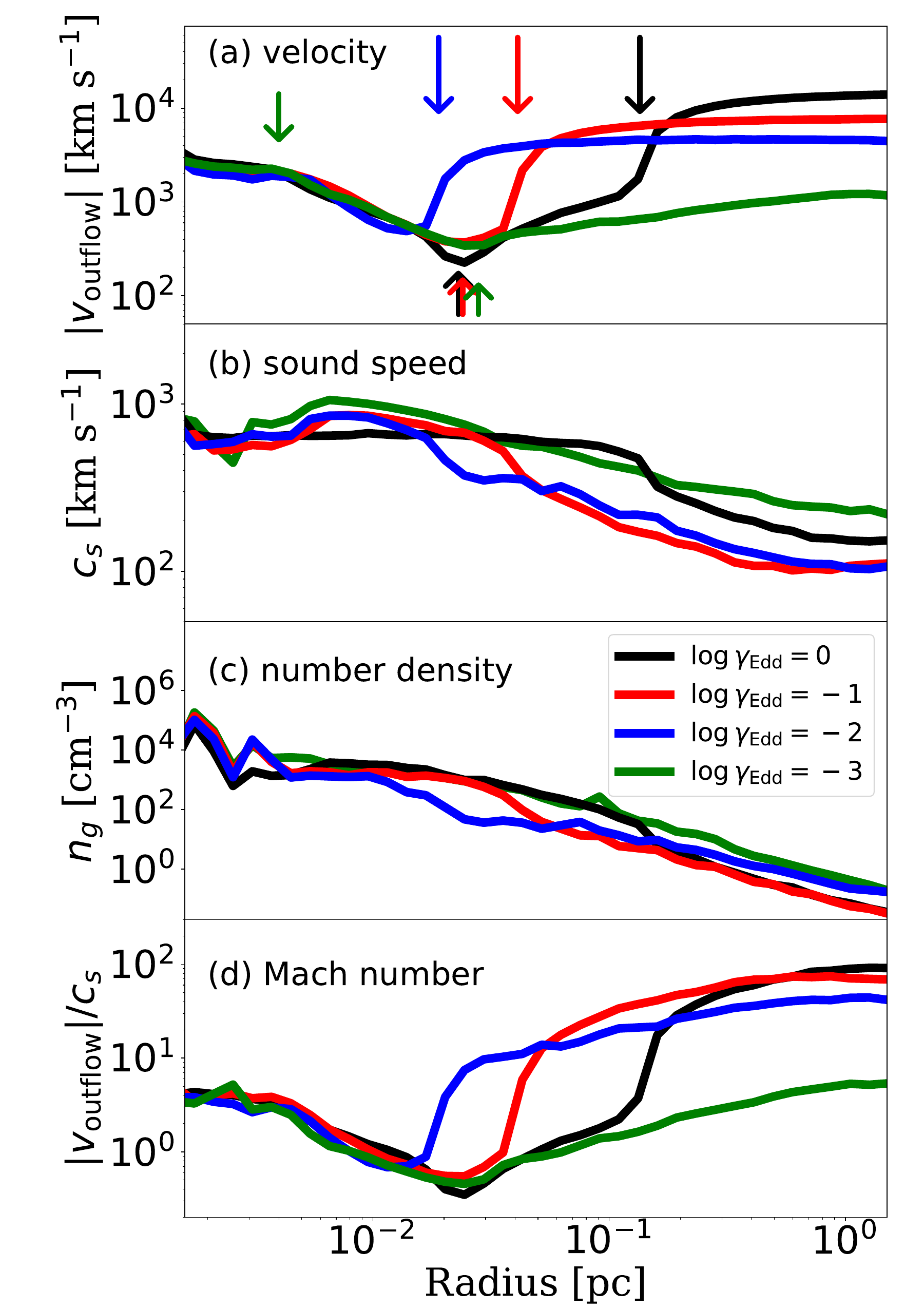}
%\end{interactive}
\caption{
Same as Figure \ref{fig:4}, but showing the radial profiles of $\log \left(\gamma_{\rm Edd} \right)= 0$ (black), $-1$(red), $-2$(blue), and $-3$(green).
In the panel (a), downward arrows indicate the dust sublimation radius, while upward arrows denote the radius where the velocity reaches a minimum at a characteristic radius.
}
\label{fig:5}
\end{figure}
%%%%%%%%%%%%%%%%%%%%

Figure \ref{fig:5} presents a comparison of the radial profiles at 60$^{\circ}$ for various Eddington ratios.
In Figure \ref{fig:5}a, the velocity structure exhibits three distinct features.
First, except for $\log (\gamma_{\rm Edd})= -2$, there is a minimum obtained at approximately 0.01 pc, indicated by the upper side arrows.
Second, except for $\log (\gamma_{\rm Edd})= -3$, there is a sharp increase in velocity by nearly an order of magnitude at certain radii, indicated by the lower side arrows.
Finally, the velocities are limited to a constant value for large radii.
The sound speed and density in Figures \ref{fig:5}b and \ref{fig:5}c indicate their minimal dependence on the Eddington ratio; however, there is a slight decrease at the radii indicated by the lower side arrows in Figure 5a.
The Mach number in Figure \ref{fig:5}d reflects the outflow velocity in Figure \ref{fig:5}a.
The outflow transitions from supersonic to subsonic after passing through the minimum point indicated by the upper side arrows in Figure \ref{fig:5}a, and then returns to supersonic.
The physics behind the structure of these velocity profiles is clarified in the following section.

\section{Analytical solutions of the AGN dusty wind}  \label{sec:analytical}
%%%%%%%%%%%%%%%%%%%%
\begin{figure}[h!]
%\begin{interactive}{animation}{fig4.mp4}
\epsscale{0.5}
\plotone{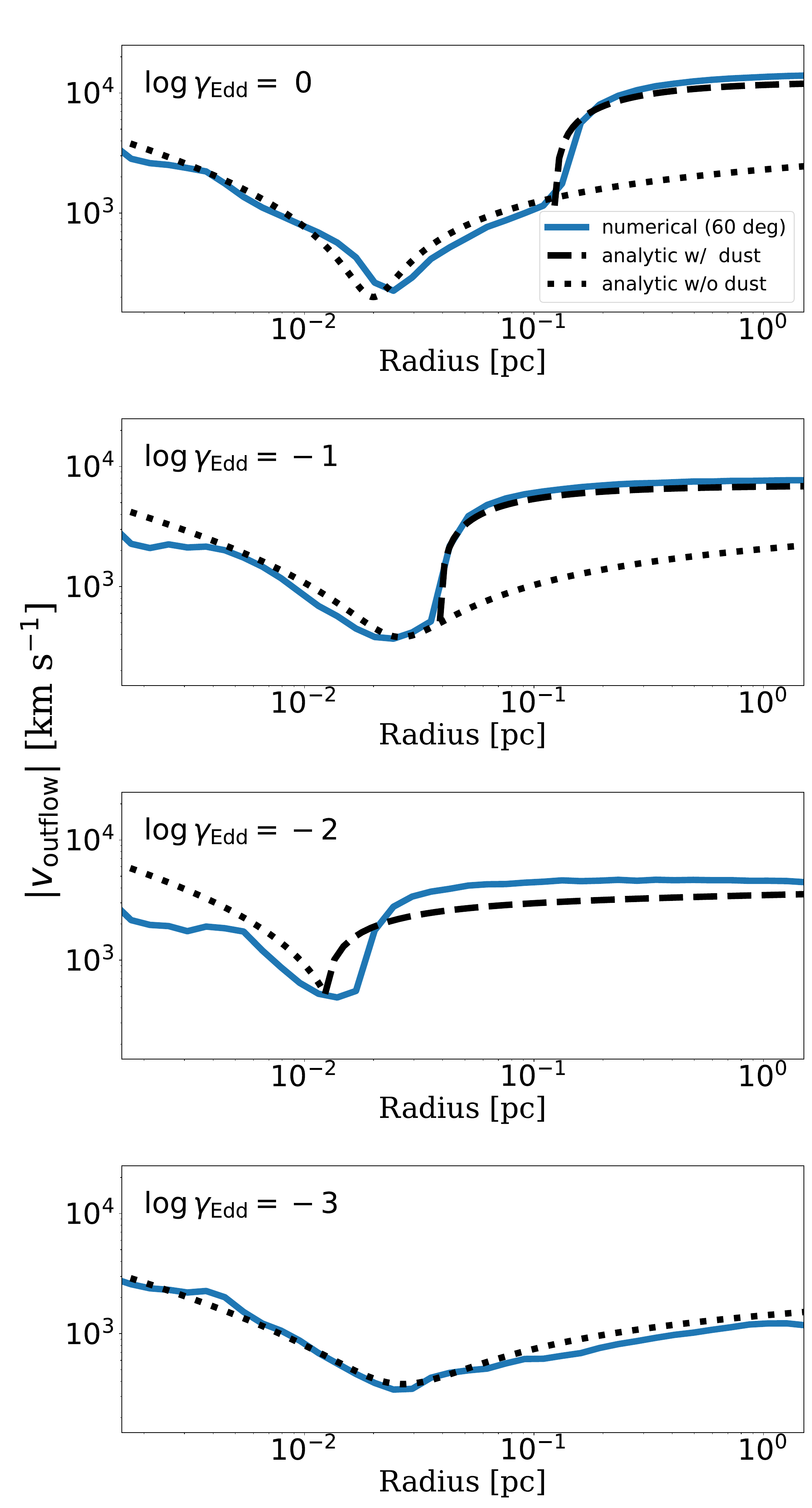}
%\end{interactive}
\caption{
Comparison between analytical solutions  (black dashed and dotted lines) and simulation results (blue solid lines) for the outflow velocity.
The blue lines correspond to Figure \ref{fig:4}c.
Dotted lines indicate the analytical solution without dust-radiation force, while dashed lines represent the solution with dust-radiation force.  
\label{fig:6}}
\end{figure}
%%%%%%%%%%%%%%%%%%%%

In this section, we present an analytical model detailing the radial profile of the ionized dusty outflow in AGNs. 
Figure \ref{fig:6} presents a comparison of the numerical results and the analytic solutions explained in \S 4.1 and 4.2.
It shows the radial distribution of the time-averaged outflow velocity at $\theta = 60^{\circ}$, plotted by the blue solid lines, similar to that in Figure \ref{fig:5}a. 
The dotted and dashed lines represent the profiles of the driving force for the gas pressure and the dust-radiation force, respectively. 
Here, we consider the time-averaged profile as representative of the outflowing gas. 
Our objective is to describe the influence of the Eddington ratio on various radii that determine the dynamical structure of the winds.

\subsection{Radial profile of outflow velocity}\label{sec:sub41}

Let us consider the time-averaged position of the shells, $r(t)$, as a function of time. 
The velocity is defined as $v(t)=dr(t)/dt$. 
By examining the forces acting on the shell, as expressed in Equation (\ref{eq:momentum}), and considering isothermal conditions and $d/dr \sim r^{-1}$, we derive the following equation for the velocity profile along the line of sight:
\begin{equation}
\frac{dv(r)}{dt} \sim \left( \Gamma_{\rm Edd}-1 \right) \frac{ r_{\rm S} }{r} \frac{c^2}{2 r} + \frac{c_s^2}{\gamma r}, 
\label{eq:ana_eom}
\end{equation} 
where $c$ and $c_s$ are the speed of light and sound, respectively. 
Further, $ r_{\rm S} $ is the Schwarzschild radius, defined as $ r_{\rm S}  = 2 G M_{\rm SMBH} /c^2 \sim10^{-6} \text{ pc } \left( M_{\rm SMBH}/10^7 M_{\odot} \right)$.
$\Gamma_{\rm Edd}$ is defined as the ratio of radiation and gravity forces,
\begin{equation}
\Gamma_{\rm Edd}=\frac{ \delta_L \kappa L_{\rm bol} }{4 \pi c GM_{\rm SMBH}}.
\end{equation}
To distinguish between dusty and dust-free gases, we add ``d" and ``g" to the index to represent the ratio of each gas,
\begin{equation}
 \Gamma_{\rm Edd, d} = \frac{\kappa }{ \kappa_{\rm T} } \delta_L \gamma_{\rm Edd},   
  \text{ and }   
\Gamma_{\rm Edd, g} =\delta_L \gamma_{\rm Edd},
\label{eq:gamma_dg}
\end{equation}
where $\Gamma_{\rm Edd, g}$ represents the case of Thomson scattering for the dust-free gas, that is $\kappa= \kappa_{\rm T}$. 
Further, $\Gamma_{\rm Edd, d}$ denotes the dusty gas with opacity expressed as $(\delta_{\rm dg} \kappa_{\rm d} + \kappa_{\rm T})$, where the dust-to-gas mass ratio is $\delta_{\rm dg}=0.01$. 
Our opacity model yielded $(\delta_{\rm dg} \kappa_{\rm d} + \kappa_{\rm T})/\kappa_{\rm T} \sim 480$. Here, $\delta_L$ is a conversion factor, for example $L_{\rm UV} = \delta_L L_{\rm bol}$, with a value of $\delta_L = 0.5$ at $\theta=60^{\circ}$.
If $\delta_{\rm L}=1$ and $\kappa =\kappa_{\rm T}$, then $\Gamma_{\rm Edd, g}= \gamma_{\rm Edd}$.

The radial velocity profile is derived by integrating Equation \ref{eq:ana_eom} from $r_0$ to $r$:
\begin{equation}
\left( \frac{v(r)}{v_0} \right)^2 = 1 + \left( \Gamma_{\rm Edd} - 1 \right) \left( \frac{ r_{\rm S} }{r_0} \right) \left( \frac{c}{v_0} \right)^2 \left[ 1 - \left( \frac{r}{r_0} \right)^{-1} \right] + \frac{1}{\gamma} \left( \frac{c_s}{v_0} \right)^2 \ln \left( \frac{r}{r_0} \right),
\label{eq:ana_solution}
\end{equation}
where $v_0$ is the reference velocity at the reference radius $r_0$. 
The radiation force acting on the outflow indicates the sign of the factor $(\Gamma_{\rm Edd}-1)$ in Equation \ref{eq:ana_eom}.

Referring to Figure \ref{fig:2}, we showed that the radiation-driven and thermal winds switched at the Eddington ratio.
The typical radius in the radiation-driven wind is the dust sublimation radius.
Equation \ref{eq:r_sub} is replaced as a function of Eddington ratio,
\begin{equation}
r_{\rm sub} =\left( \delta_L  \gamma_{\rm Edd} \right)^{\frac{1}{2}}  r_{\rm sub, Edd },
\label{eq:r_sub_fgamma}
\end{equation}
where $r_{\rm sub, Edd }$ is the value normalized to the Eddington luminosity $L_{\rm Edd}$,
\begin{equation}
\quad  r_{\rm sub, Edd } =  \sqrt{ \frac{ L_{\rm Edd}}{4 \pi \sigma_{\rm SB}  T_d^{4} } }
\sim 0.176 \text{ pc } \left( \frac{M_{\rm SMBH} }{10^7 M_{\odot}} \right)^{1/2}  \left(\frac{T_{\rm d}}{1500 \text{ K } } \right)^{-2}. 
\label{eq:r_sub_Edd}
\end{equation}
The thermal wind can have the critical point $r_{\rm cr}$, following from the equilibrium condition in Equation \ref{eq:ana_eom},
\begin{equation}
r_{\rm cr} =\left(1- \Gamma_{\rm Edd} \right) \frac{\gamma}{2} \left( \frac{c}{c_s} \right)^2 r_{\rm S}. 
\label{eq:r_cr}
\end{equation}
Similar to Equation \ref{eq:gamma_dg}, we use "d" and "g" indices for dusty and dust-free gases, denoted as  $r_{\rm cr, d}$ and $r_{\rm cr, g}$, respectively.
This radius is a monotonically decreasing function for $\gamma_{\rm Edd}$.
When $\Gamma_{\rm Edd} \rightarrow 0$, the critical radius has the maximum, that is, $r_{\rm cr} = 0.017 \text{ pc } \left( (c/c_s)/200 \right)^2 \left( M_{\rm SMBH}/10^7 M_{\odot} \right) $.
When $\Gamma_{\rm Edd, d} \rightarrow 1$ for the dusty gas, then $r_{\rm cr, d}= 0$ and a similar result for the dust-free gas. 
There is no critical point when $\Gamma_{\rm Edd, g}>1$ for the dust-free gas or $\Gamma_{\rm Edd, d}>1$ for the dusty gas.

Figure \ref{fig:6} presents a comparison of the simulation results (blue lines) with the analytic solutions (black lines) derived using Equation \ref{eq:ana_solution}. 
By incorporating the outflow velocity at $r=r_{\rm sub}$ into Equation \ref{eq:ana_solution} as $r_0$ and $v_0$, we can reproduce the profiles of the outflow velocity. 
In the case of $\log \gamma_{\text{Edd}} = 0$ in the upper panel of Figure \ref{fig:6}, there are three domains of solutions (see also the top panel of Figure \ref{fig:7}).
(a) For $r < r_{\text{cr}}$, the wind decreases monotonically owing to the force of gravity (dotted line). 
(b) For $r_{\text{cr}} < r < r_{\text{sub}}$, the gas pressure overcomes the gravity, accelerating the outflow. 
(c) For $r > r_{\text{sub}}$, the outflow path switches from the dotted line to the dashed line, and the radiation force becomes dominant. 
The solutions without dust-radiation force, indicated by the dotted lines, are similar to the canonical model of supersonic solutions in solar/stellar winds \citep{1960ApJ...132..821P}. 
Furthermore, the solution with dust-radiation force, indicated by the dashed lines, corresponds to the solutions for stellar dusty winds \citep{1999isw..book.....L}.

For $\log \gamma_{\rm Edd}=-2$, there is no critical point, and only the solution (c) is obtained. 
Conversely, when $\log \gamma_{\rm Edd}=-3$, following solution (a), acceleration owing to gas pressure continues through the critical point $r_{\rm cr, d}$.
Notably, despite passing through the dust sublimation radius, solution (c) does not occur due to the weaker dust-radiation force compared to the gravity, as described by Equation \ref{eq:gamma_dg}, that is $\Gamma_{\rm Edd, d} = 0.24 \left( (\kappa/\kappa_{\rm T})/ 480 \right) \left( \delta_L / 0.5 \right) (\gamma_{\rm Edd}/10^{-3} ) < 1$.

\subsection{Terminal velocity}\label{sec:sub42}

The outflow velocity in Figure \ref{fig:6} tends to reach a terminal velocity in the outer region (i.e., $r \sim 1$ pc) for $\log \gamma_{\rm Edd} \gtrsim -2$, as referenced in Figure \ref{fig:5}a.
The terminal velocity can be derived by energy conservation, similar to the process for radiation-driven stellar winds. 
By equating the kinetic energy at $r \rightarrow \infty$ with the total energy at $r=r_{\rm sub}$, the terminal velocity $v_{\infty}$ is expressed as,
\begin{equation}
\left( \frac{v_{\infty}}{v (r_{\rm sub})} \right)^2 = 1+  \left( \Gamma_{\rm d, Edd}-1 \right) \left( \frac{c}{v(r_{\rm sub})} \right)^2 \frac{ r_{\rm S} }{r_{\rm sub}} + \frac{2}{\left( \gamma-1 \right) \gamma} {\cal M}(r_{\rm sub})^{-2},
\label{eq:terminal_v}
\end{equation}
where $\cal M$ is the Mach number.
The dominant term in Equation \ref{eq:terminal_v} is the second term of the dust-radiation force on the right-hand side,
\begin{equation}
\begin{split}
 \frac{v_{\infty}}{v (r_{\rm sub})}  
&\sim
 \Gamma_{\rm Edd, d}^{1/2} \frac{c}{v(r_{\rm sub}) }   \left(\frac{ r_{\rm S} }{r_{\rm sub}} \right)^{1/2} \\ 
&= \sqrt{ \frac{\kappa }{\kappa_{\rm T}} \frac{ r_{\rm S} }{r_{\rm sub, Edd} }  \left( \frac{c}{v(r_{\rm sub})} \right)^2 } \left( \delta_L \gamma_{\rm Edd} \right)^{1/4},
\end{split}
\label{eq:terminal_eddington}
\end{equation} 
where $ r_{\rm S}  / r_{\rm sub, Edd} \propto M_{\rm SMBH}^{1/2}$ according to Equation \ref{eq:r_sub_Edd}.
We found that $v_{\infty}$ scales with $\gamma_{\rm Edd}^{1/4}$ and $M_{\rm SMBH}^{1/4}$, determined at $r=r_{\rm sub}$.
In Figure \ref{fig:5}a, the terminal velocity is $v_{\infty} /v (r_{\rm sub}) \lesssim 10$. 
This is consistent with the estimation using Equation \ref{eq:terminal_eddington}, 
$
v_{\infty} / v (r_{\rm sub}) \sim 11 \left( c /v(r_{\rm sub}) / 250 \right) \left( M_{\rm SMBH}/ 10^7 M_{\odot} \right)^{1/4} \left( \gamma_{\rm Edd} / 1 \right)^{1/4}
$
.

\subsection{Eddington ratio dependence in the analytic solution}\label{sec:sub43} 

%%%%%%%%%%%%%%%%%%%%
\begin{figure*}[h!]
%\begin{interactive}{animation}{fig4.mp4}
\epsscale{1.0}
\plotone{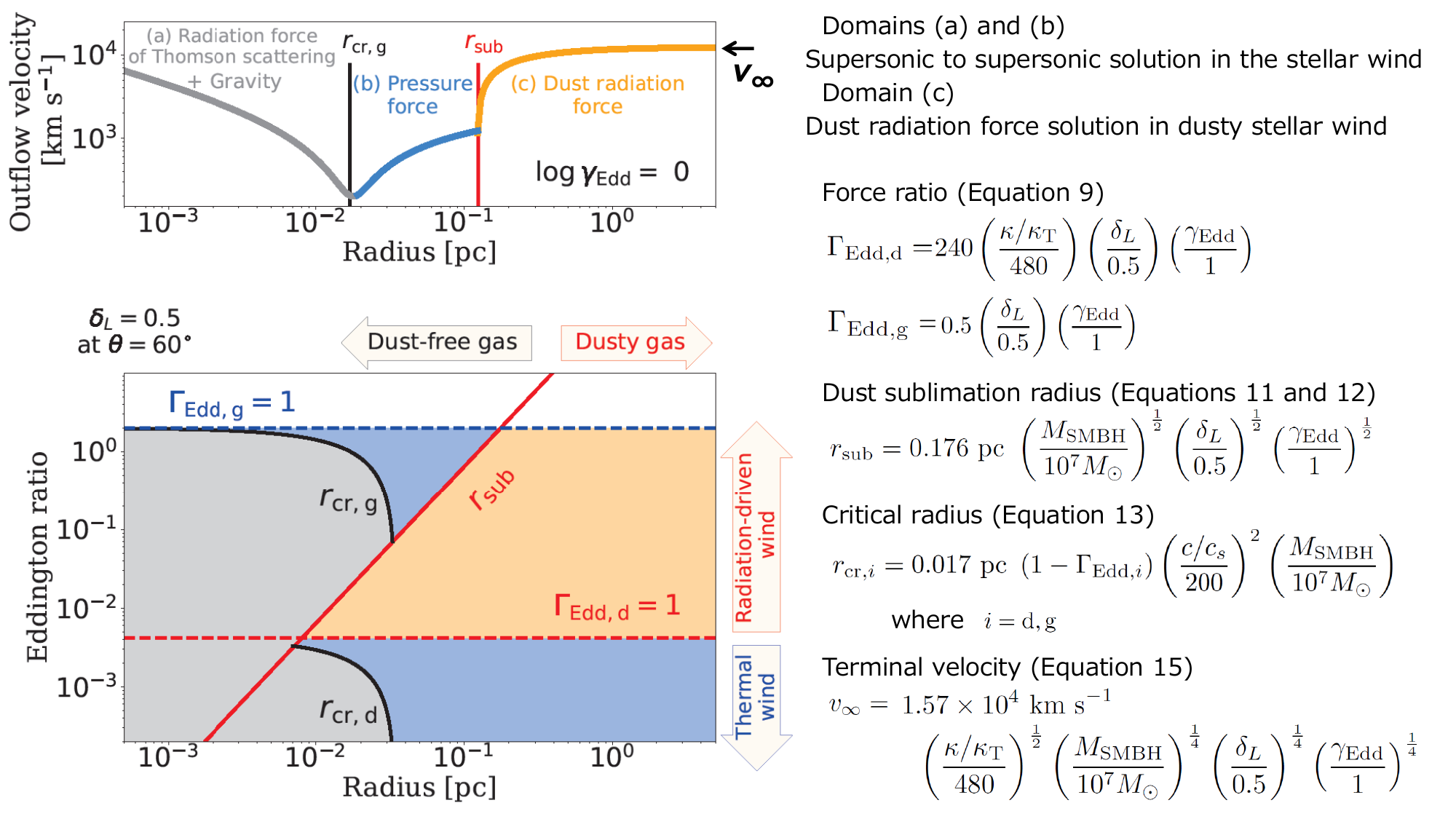}
%\end{interactive}
\caption{
Top panel is a typical radial profile of the outflow based on the analytical solution for $\log\gamma_{\rm Edd}=0$.
Bottom panel indicates the region of the dominated force and the typical radii on the $\log r$--$\log \gamma_{\rm Edd}$ plane.
Three dominant forces are delineated: acceleration by gas pressure (blue) and by dust-radiation force (orange), and deceleration by gravity (gray).
These regions are distinguished by the dust sublimation radius (red solid line) and the critical point (black solid line).
The dust-radiation force is also dependent on $\Gamma_{\rm Edd, d}$ (dashed line), representing the optically thin case of \citet{2008MNRAS.385L..43F}.
The right-hand side summarizes the processes and Equations related as the analytic solution. 
\label{fig:7}}
\end{figure*}
%%%%%%%%%%%%%%%%%%%%

Figure \ref{fig:7} summarizes the analytical solutions explained in 
\S \ref{sec:sub41} and \ref{sec:sub42} for $\theta=60^{\circ}$, corresponding to the optically thin situation (i.e., $N_H \sim 10^{22}$ cm$^{-2}$).
In the upper panel, the dynamic structure is explained by the transition of dominant forces at two typical radii.
Initially, the outflow injected by radiation energy propagates while decelerated by gravity until it reaches the critical radius $r_{\rm cr}$ (domain a).
Subsequently, gas-pressure acceleration occurs (domain b).
Beyond the dust sublimation radius $r_{\rm sub}$, the dust-radiation force becomes dominant, causing the outflow to rapidly accelerate towards its terminal velocity (domain c).

In the lower panel of Figure \ref{fig:7}, the colored regions show the structures of the outflows on the $\log r$--$\log \gamma_{\rm Edd}$ plane in the same color scheme as that used in the upper panel.
The solid red line represents the dust sublimation radius $r_{\rm sub}$, dividing the region into dusty and dust-free gases. 
This radius, described in Equation \ref{eq:r_sub_fgamma} 
\footnote{
Equation \ref{eq:r_sub} provides the value under the assumption of the optically thin limit. 
However, the actual dust sublimation radius, $r_{\rm sub}$, is influenced by factors such as anisotropic radiation, as depicted in Figure \ref{fig:1}b, and attenuation owing to optical depth $\tau$. 
\citetalias{2023ApJ...950...72K} showed that for Thomson scattering, the dust sublimation radius decreases rapidly under optically thick conditions ($\tau > 1$), as indicated by the black contours in Figures \ref{fig:2} and \ref{fig:3}.
 }
, increases in proportion to $\gamma_{\rm Edd}^{1/2}$.
The orange region indicates the dominance of the dust-radiation force over other forces, occurring when $r>r_{\rm sub}$ (red solid line) and $\Gamma_{\rm Edd, d}>1$ (red dashed line).
The condition $\Gamma_{\rm Edd, d}>1$ from Equation \ref{eq:gamma_dg} signifies that the dust-radiation force surpasses the gravity.
$\Gamma_{\rm Edd, d}$ becomes larger than $\gamma_{\rm Edd}$ and proportional to the opacity ratio between the dusty and dust-free gases
\footnote{
This condition means super Eddington for the dust-radiation force \citep{2008MNRAS.385L..43F}.
When the optically thick extreme situation is considered, $\Gamma_{\rm Edd}$ shifts to larger $\gamma_{\rm Edd}$ because of $\delta_L$ effects, i.e. the attenuation and radiation anisotropy.
This is similar to a function on the plane of column density and $\gamma_{\rm Edd}$ \citep{2018MNRAS.479.3335I}. 
$\Gamma_{\rm Edd} \sim 10^{-3} \gamma_{\rm Edd}$ is valid for the optically thin case.
}.
The $\Gamma_{\rm Edd, d}=1$ distinguishes the two driving mechanisms of the outflow (see Figure \ref{fig:2}).

The solid black lines represent the critical radii, denoted as $r_{\rm cr,g}$ and $r_{\rm cr, d}$, in the regions of dust-free and dusty gases, respectively.
The blue region indicates the dominance of the gas pressure over other forces, occurring when $r_{\rm cr, g} < r < r_{\rm sub}$ for dust-free gas and $r>r_{\rm cr, d}$ for dusty gas. 
In situations where the opacity is subjected to a discontinuous change at $r_{\rm sub}$, $r_{\rm cr,g}$ and $r_{\rm cr, d}$ do not coexist and are limited to $r_{\rm sub}$.
According to Equation \ref{eq:r_cr}, $r_{\rm cr}$ is the decreasing function proportional to $(1- \Gamma_{\rm Edd})$ and $M_{\rm SMBH}$ with respect to $\gamma_{\rm Edd}$.
Further, $r_{\rm cr, g}$ and $r_{\rm cr, d}$ cannot exist for $\Gamma_{\rm Edd, g}>1$ (blue dashed line) and $\Gamma_{\rm d, Edd}>1$ (red dashed line).
Although $r_{\rm cr, g}$ exists at $\log \gamma_{\rm Edd}=0,$ and $-1$, $r_{\rm cr, d}$ exists at $\log \gamma_{\rm Edd}=-3$, and no critical points exist at $\log \gamma_{\rm Edd}=-2$.
These results are consistent with our simulation results (also see Figure \ref{fig:6}).

\section{Discussion}\label{sec:discussion}

\subsection{Applications of the analytic solutions} \label{sec:sub51}

The proposed analytical model facilitates a new interpretation that is distinct from conventional wind studies.
Recent investigations into wind dynamics with the self-similar solutions have used two approaches: the shock propagation and steady models. 
The former addresses the propagation of a single shocked shell \citep{2012MNRAS.425..605F,2015ARA&A..53..115K,2018MNRAS.473.4197C,2020MNRAS.497.5229C,2024MNRAS.528.6496H}. 
In contrast, the latter focuses on steady solutions for radiation forces, including continuum \citep{2017ApJ...846..154C}, infrared \citep{2016ApJ...819..115D,2020ApJ...900..174V}, and combinations with magnetocentrifugal force \citep{2005ApJ...631..689E}. 
\citet{2012MNRAS.426.2239W,2019MNRAS.489.1152M,2020ApJ...890...80C} demonstrated the disk winds driven by gas pressure. 
In our analytical model, we confirm that the forces driving winds on the plane of Eddington ratio and radius are determined by $\Gamma_{\rm Edd}$, $r_{\rm cr}$, and $r_{\rm sub}$. 
Despite the repeated occurrence of shocks in the simulation results, demonstrating such behavior as an analytical solution derived from the steady wind represents a new interpretation. This can be applied to a statistical understanding or to an average outflow processes of individual objects.

The proposed analytical solution provides a method for predicting the velocity of dust-free gas within the dust sublimation radius based on the observations of ionized dusty outflow at the parsec scale. 
We outline a procedure for determining the dynamical structure along the line of sight on the ionized dusty outflows. 
By assuming that the observed velocity ($v_{\rm obs}$) becomes terminal at a spatial scale ($r_{\rm obs}$) on the parsec scale, we apply these values to $v_0=v_{\rm obs}$ and $r_0=r_{\rm obs}$ in Equation \ref{eq:ana_solution}.
Considering the values of $r_{\rm sub}$ from Equation \ref{eq:r_sub}, determined based on the Eddington ratio and SMBH mass, a solution for the radiation-driven wind using $\Gamma_{\rm Edd, d}$ (Equation \ref{eq:gamma_dg}) down to $r=r_{\rm sub}$ can be obtained using Equation \ref{eq:ana_solution}. 
For $r < r_{\rm sub}$, it is set for $v_0=v(r_{\rm sub})$ and $r_0=r_{\rm sub}$ in the solution of the radiation-driven wind.
Similarly, a solution for dust-free gas be obtained using Equation \ref{eq:ana_solution} employing $\Gamma_{\rm Edd, g}$ (Equation \ref{eq:gamma_dg}).
Our model incorporates the conversion factor $\delta_L$ for the bolometric and UV luminosities. 
The UV luminosity represents the $L_{\rm corona}$ and the anisotropy of $L_{\rm AD}$ (see also Figure \ref{fig:1}), depending on the inclination angle of an object.
However, the radiation force can decrease owing to the attenuation of the column density. 
The column density evaluated by X-ray observation provides the optical depth, $\tau \sim (\kappa_{T}/0.41)(N_H/10^{24} \text{ cm}^{-2})$. 
Therefore, the UV luminosity decreases as $\delta_L e^{-\tau}$.
Considering the angular dependence of column density, as observationally discussed by \citet{2023ApJ...959...27R}, facilitates the discussion of the radial distribution at different angles.

The terminal velocity in our analytical model is reached at a position 10 times the dust sublimation radius (Figure \ref{fig:5}a, excluding green line). 
As shown in Equation \ref{eq:terminal_eddington}, the terminal velocity is dependent on $M_{\rm SMBH}^{1/4}$ and $\gamma_{\rm Edd}^{1/4}$. 
In other words, it predicts the scaling $ L_{\rm bol}^{1/4}$,
\begin{equation}
v_{\infty}  \sim 1.1 \times 10^4 \text{ km s}^{-1} \quad
\left( \frac{\delta_L }{ 0.5} \right)^{1/4}
\left( \frac{\kappa/\kappa_T}{480} \right)^{1/4}
\left( \frac{L_{\rm bol}}{ 10^{45} \text{ erg s}^{-1} } \right)^{1/4}.
\label{eq:v_inf_dis}
\end{equation}
\citet{2017A&A...601A.143F} identified the correlation with the bolometric luminosity for the multi-phase outflow, scaling of $v_{\rm obs} \propto L_{\rm bol}^{1/5}$ on the $v_{\rm obs}$--$L_{\rm bol}$ plane. 
Compared to their results, the magnitude of the velocity in Equation \ref{eq:v_inf_dis} is more indicative of an X-ray wind (warm absorber) than ionized outflow. 
\citet{2022ApJ...925...55O} verified the absorption feature of the X-ray wind in the parsec-scale ionized gas using simulations from \citet{2016ApJ...828L..19W}.

Our solution suggests that the launch scale of ionized dusty outflows may provide insights into the BLR scale. 
The angle for Compton-thin implies a dust-free BLR \citep[][]{2016MNRAS.462.3570S,2022ApJS..261....8R,2023ApJ...943...63N}. 
At angles corresponding to the Compton thick, the dust sublimation radius decreases rapidly due to the outflow's own optical thickness. 
As dust can survive inside $r_{\rm sub, Edd}$ (Equation \ref{eq:r_sub_Edd}), a dusty BLR would be formed on the disk surface \cite[][]{2011A&A...525L...8C,2023EPJD...77...56C,2024Univ...10...29N}. 
The optical thickness related to Thomson scattering is discussed as column density in the following section.

\subsection{Column density } \label{sec:sub52}
%%%%%%%%%%%%%%%%%%%%
\begin{figure}[h!]
%\begin{interactive}{animation}{fig4.mp4}
\epsscale{0.5}
\plotone{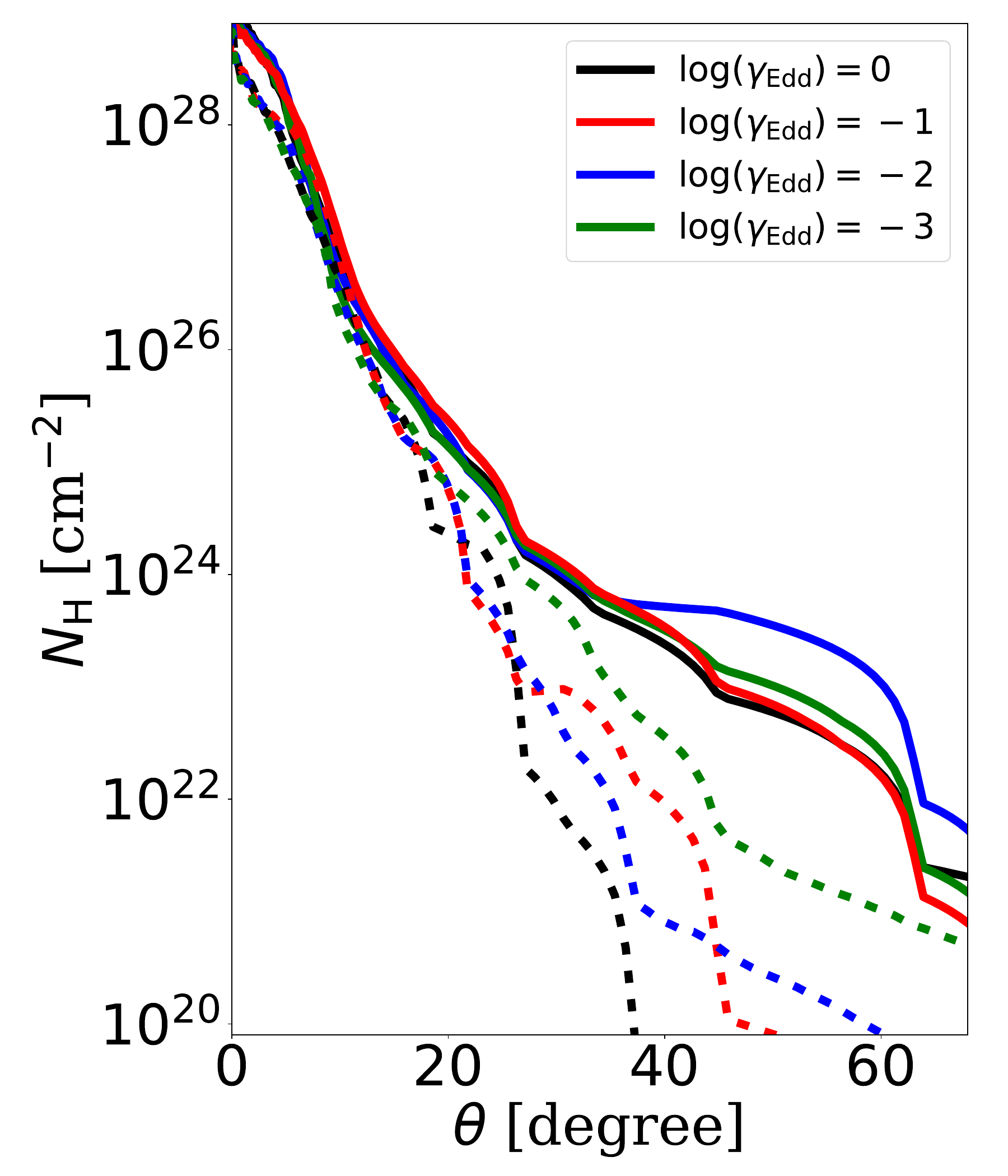}
%\end{interactive}
\caption{
Time-averaged column density $N$ as a function of $\theta$.
Solid lines denote both the dusty and dust-free component, while dotted lines indicate the dusty component. 
The colors correspond as $\log (\gamma_{\rm Edd})=0$(black), $-1$(red), $-2$(blue), and $-3$(green).
\label{fig:8}}
\end{figure}
%%%%%%%%%%%%%%%%%%%%

Figure \ref{fig:8} shows the column densities as a function of $\theta$ for the dusty and dust-free gases.
The larger the Eddington ratio, the smaller the column density for the dusty gas (dotted lines).
Clearly, the dust-free gas (solid lines) dominates the column density at all angles, and is independent of the Eddington ratio.

Using the mass conservation, $n(r)v(r)r^2=\text{const.}$, the column density shown in Figure \ref{fig:8} can be estimated from the analytical solution provided in Equation \ref{eq:ana_solution}.
This estimation achieves integration from $r_{\rm sub}$ to a certain radius $r_{\rm e}$,
\begin{equation}
N_H =  \int_{r_{\rm sub}}^{r_{\rm e}} n(r_{\rm sub}) \left( \frac{v(r_{\rm sub})}{v(r)} \right) \left( \frac{r_{\rm sub}}{r} \right)^2 dr.
\label{eq:column}
\end{equation}
We define the the normalized value relative to the dust sublimation scale as,
\begin{equation}
N_{H, {\rm sub}}  = n(r_{\rm sub}) r_{\rm sub} \left( \frac{ r_{\rm sub} }{ r_{ g} } \right)^{1/2} \left( \frac{v (r_{\rm sub}) }{c} \right)
\sim 10^{20} \quad \text{cm}^{-2},
\label{eq:column_sub}
\end{equation}
which is derived from the gravity term in Equation \ref{eq:ana_solution}.
This formulation provides a straightforward understanding of ionized dusty outflow for the dusty and dust-free gases.

For the case of the radiation-driven wind, the largest contribution on the right-hand side of Equation \ref{eq:ana_solution} is the radiation force term.
Thus, Equation \ref{eq:column} yields,
\begin{align}
N_H \sim N_{H, {\rm sub}} \times
\begin{cases}
\Gamma_{\rm Edd, d}{}^{-1/2} 
& 
\quad \text{for dusty gas},  
\\
\left( \frac{r_{\rm inj}}{r_{\rm sub}}\right)^{-1/2}
& 
\quad  \text{for dust-free gas}.
\end{cases}
\label{eq:column_order}
\end{align}
The column density for the dusty gas is determined by integrating the dust-radiation-force term, including $\Gamma_{\rm Edd, d}$, as $r_{\rm e} \rightarrow \infty$. 
The column density for the dusty gas exhibits a decreasing trend with the Eddington ratio.
However, the column density for the dust-free gas is determined by integrating the gravity force term to the injection scale $r_{\rm e}= r_{\rm inj}$.
Assuming $ r_{\rm inj}/r_{\rm sub} \sim 10^{-4}$, $N_H \sim 10^{22}$ cm$^{-2}$ is achieved, which approximately explains the solid line in Figure \ref{fig:8}. 
Note that Equation \ref{eq:column_order} yields a lower limit as the analytical solution does not consider the launch scale from the disk to the outflow.
In the case of $\log \gamma_{\rm Edd}=-3$, the thermal wind is independent of the Eddington ratio because the dominant contribution is attributed to the gas pressure force in the third term of Equation \ref{eq:ana_solution}.
These analytical findings approximately correspond to the numerical results of $\theta=60^{\circ}$ and its behavior in Figure \ref{fig:8}.

\subsection{Obscuring fraction} \label{sec:sub53}

%%%%%%%%%%%%%%%%%%%%%%%%%%
\begin{figure*}[h!]
%\begin{interactive}{animation}{fig4.mp4}
\epsscale{1.2}
\plotone{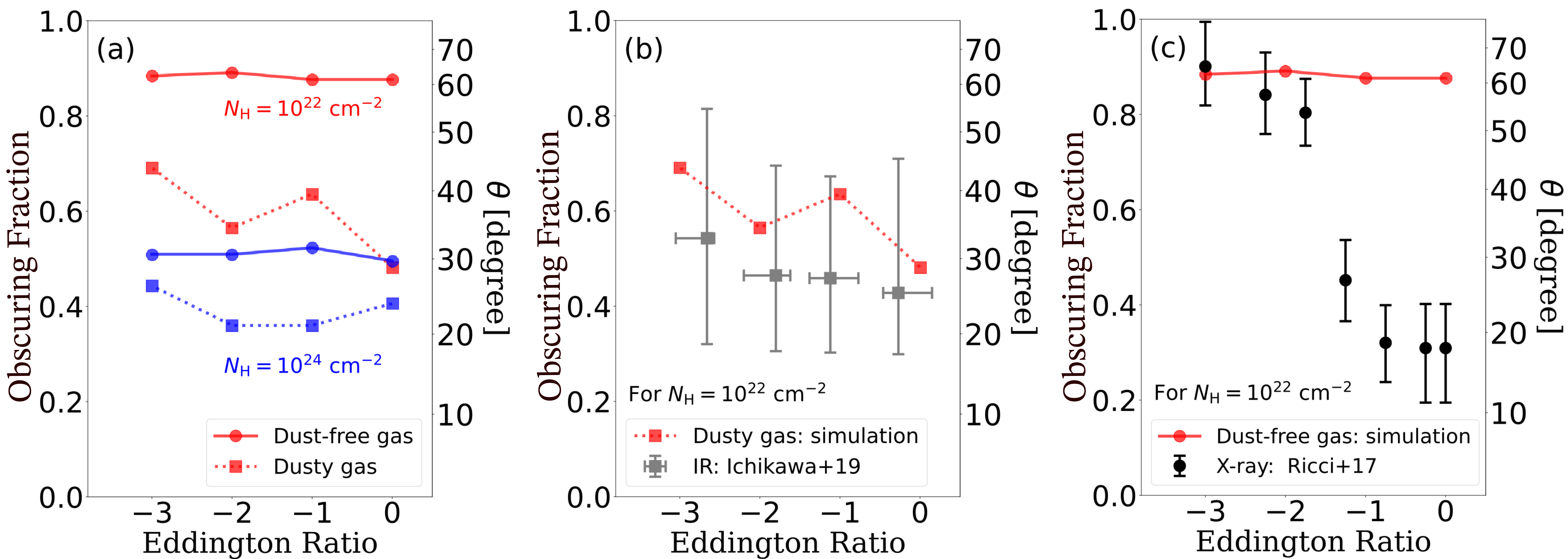}
%\end{interactive}
\caption{
(a) Obscuring fraction $f_{\rm obs}$ of the numerical results as a function of 
the Eddington ratio for $N_H=10^{24}$ cm$^{-2}$ (blue) and $N_H=10^{22}$ cm$^{-2}$ (red).
Circle symbols with solid lines represent the dust-free gas, and square symbols with dotted lines represent the dusty gas. 
(b) Comparison of the {obscuring fraction} between the numerical results for the dusty gas (red) for $N_H=10^{22}$ cm$^{-2}$ and observations: Gray square symbols with bars represent  the median with $\pm 1 \sigma$ percentile of the observed IR sources combining the obscuring fraction $f_{\rm obs, IR}$ \citet{2019ApJ...870...31I}.
The Eddington ratios for each object are based on the results by \cite{2021ApJS..252...29K}.
(c) Same as (b), but for the dust free gas. The black circle symbols with bars represent the observed X-ray data \citep{2017Natur.549..488R}.
\label{fig:9}
}
\end{figure*}
%%%%%%%%%%%%%%%%%%%%%%%%

We discuss the fraction obscured by sub-parsec-scale outflow.
The obscuring fraction is defined as the ratio of the obscured solid angles to the hemisphere, i.e. 
$
f_{\rm obs}=2 \pi \int_0^{\theta(N_H)} \cos \vartheta d\vartheta /2 \pi =  \sin \theta (N_H)
$
In Figure \ref{fig:9}a, we show the $f_{\rm obs}$ dependence on the Eddington ratio for $N_H=10^{24}$ cm$^{-2}$ (blue) and $10^{22}$ cm$^{-2}$ (red), with respect to Figure \ref{fig:8}.
The $f_{\rm obs}$ remains nearly constant for both dusty (square symbols) and dust-free (circle symbols) gases, independent of the Eddington ratio.
Referring to Figure \ref{fig:4}a, the $f_{\rm obs}$ appears to be determined by the density surrounding the dust sublimation radius.
Thus, the sub-parsec-scale outflow exhibits a significant $f_{\rm obs}$ even when the Eddington ratio is low.

Figures \ref{fig:9}b and \ref{fig:9}c show the observed $f_{\rm obs, IR}$ \citep{2019ApJ...870...31I} and $f_{\rm obs, X}$ \citep{2017Natur.549..488R}, respectively. 
The black symbols denote the $f_{\rm obs, X}$ using the nearby AGN samples detected in the Swift/BAT 70-month survey, excluding the blazer objects.
From the same catalog, \citet{2019ApJ...870...31I} analyzed the $f_{\rm obs, IR}$, which is related to the IR luminosity of the AGN component from the SED.
We adopted their data related into the Eddington ratio reported in \citet{2021ApJS..252...29K}.
The gray symbols are the median value of $ f_{\rm obs, IR}\sim 0.5$ for the dusty gas, which is consistent with the findings of \citet{2019ApJ...870...31I}.
Both figures also show the numerical results for $N_H=10^{22}$ cm$^{-2}$ for the dusty (red dotted line) and dust-free (red solid line) gases.

In Figure \ref{fig:9}b, the weak correlation observed between the $f_{\rm obs, IR}$ and the Eddington ratio aligns with our results, consistent with recent observational results \citep[e.g.,][]{2021ApJ...912...91T,2023ApJ...959...27R}. 
They discussed that the magnitude of $f_{\rm obs}$ originated from dust emitted in the polar direction or dusty outflows depending on the column density.
In simulation studies for the dusty gas, \cite{2015ApJ...812...82W} reported that the significant $f_{\rm obs}$ at a column density of $N_H=10^{22}$ cm$^{-2}$ is associated with a parsec-scale dusty torus,
which is formed by the circulation of dusty gas driven by the central radiation.
Conversely, \cite{2020ApJ...897...26W} suggested that it primarily constitutes an outflow.
Furthermore, \cite{2020ApJ...889...84K} analytically indicated that the phenomena on parsec scale, such as fountain flows and star formation, exhibit a obscuring fraction for dusty gas within 0.6.

Our results indicate that $f_{\rm obs}$ and the column density of the dusty gas are determined by the sub-parsec-scale outflow. 
From the analytical solution, the column density is determined by the Eddington ratio and the number density of the dust sublimation scale in Equation \ref{eq:column_order}, that is $N_{\rm H, sub}$ and $\Gamma_{\rm Edd, d}$ (Equations \ref{eq:gamma_dg} and \ref{eq:column_sub}). 
However, a significant variability is observed in the IR extinction within nearby AGN, indicating the involvement of multiple factors in determining the column density of dusty gas \citep{2021A&A...652A..99A,2022ApJ...940...28K,2023A&A...678A.136I}.  
Numerically, the 10 parsec-scale fountain structure presented by \citet{2015ApJ...812...82W} is related to the sub-parsec-scale dust outflow discussed by \citetalias{2023ApJ...950...72K}. 
Thus, both the dust sublimation scale owing to outflow and the accumulation on parsec scale owing to the fountain effect contribute to the dust column density. 
Therefore, we argue for caution in the interpretation of the dust-gas column density in relation to emission and extinction.

In the range of $\gamma_{\rm Edd} \lesssim -2$ shown in Figure \ref{fig:9}c, $f_{\rm obs}$ of dust-free gas (red circle symbols) appears to account for the high $f_{\rm obs, X}$ (black symbols). 
This is attributed to the presence of dust-free gas within the dust sublimation radius. 
$f_{\rm obs}$ of dust-free gas remains constant and unaffected by changes in the Eddington ratio (see also, the solid lines of Figure \ref{fig:8}). 
However, $f_{\rm obs, X}$ rapidly decreases for $\log \gamma_{\rm Edd} \gtrsim -1.5$ \citep[e.g.,][]{2017Natur.549..488R,2022ApJ...938...67R}. 
They argued that this phenomenon is a result of the gas distribution regulated by radiation. 
\citet{2022ApJS..260...30T} showed that in Compton-thick obscured AGNs, $f_{\rm obs, X}$ is larger compared to less obscured ones, but with a similar decreasing trend against the Eddington ratio.

What could be causing these discrepancies? 
In our study, we explain the column density of dust-free gas and its high $f_{\rm obs}$ through outflows induced by the radiation forces from dust and Thomson opacities, by spatially resolving the dust sublimation scale. 
However, considering the observed trend of high $f_{\rm obs, X}$ at Eddington ratios, it is necessary to consider outflow driven from within sub-parsec scales that were not considered in our simulations. 
This indicates influence from outflows originating close to the Schwarzschild radius. 
Recent studies have suggested the influence of mechanisms, e.g., magnetic-driven outflows \citep{2021ApJ...914...31Y,2022MNRAS.513.5818W} and/or line-driven outflows \citep{2019A&A...630A..94G,2021MNRAS.507..904N,2023arXiv231018557D}. 
\citet{2020MNRAS.494.3616N} demonstrated through hydrodynamic simulations at scales of $100 r_{\rm S} $ (equivalent to $\sim 10^{-4} (M_{\rm SMBH}/10^7 M_{\odot})$ pc) that the line-driven outflow is activated for $\log \gamma_{\rm Edd} \gtrsim -1.5$, with a blowout angle of $\theta\lesssim 20^{\circ}$.
Thus, considering the radiation-driven outflow launching from the Schwarzschild scale, it is expected to decrease $f_{\rm obs}$ from the perspective of blowout angles and Eddington ratios \citep{2022MNRAS.513.1141Z,2023RAA....23l5008Z}.

\subsection{Ionization state}\label{sec:sub54}

Our results for outflow velocity and its classification are likely related to the X-ray wind observed in ionized gas. 
Particularly, with respect to observed blue-shifted absorption lines,  \cite{2013MNRAS.430.1102T,2014MNRAS.441.2613L}(references after therein) have reported that the column density and outflow velocity are characterized by the ionization parameter, defined as
$ \xi=L_{\rm ion}/(nr^2) $ erg s$^{-1}$ cm
where $L_{\rm ion}$ is the $1$--$1000$ Ryd (13.6 eV -- 13.6 keV) ionizing luminosity.
\cite{2024ApJS..274....8Y} conducted a comprehensive study of the ionization dependency of X-ray outflows and revealed that they are distributed over a wide range of 
$-1 \lesssim \log \xi \lesssim 5$ and $10^2 \lesssim v_{\rm outflow} \lesssim 10^5 \text{km s}^{-1}$.
In this subsection, we discuss how the ionization parameter and its spatial distribution based on simulations and analytical solutions correspond to the observed ionization parameter.

%%%%%%%%%%%%%%%%%%%%%%%%%%
\begin{figure}[h!]
%\begin{interactive}{animation}{fig4.mp4}
\epsscale{1.0}
%\plotone{CF_NUMvsOBS.png}
\plotone{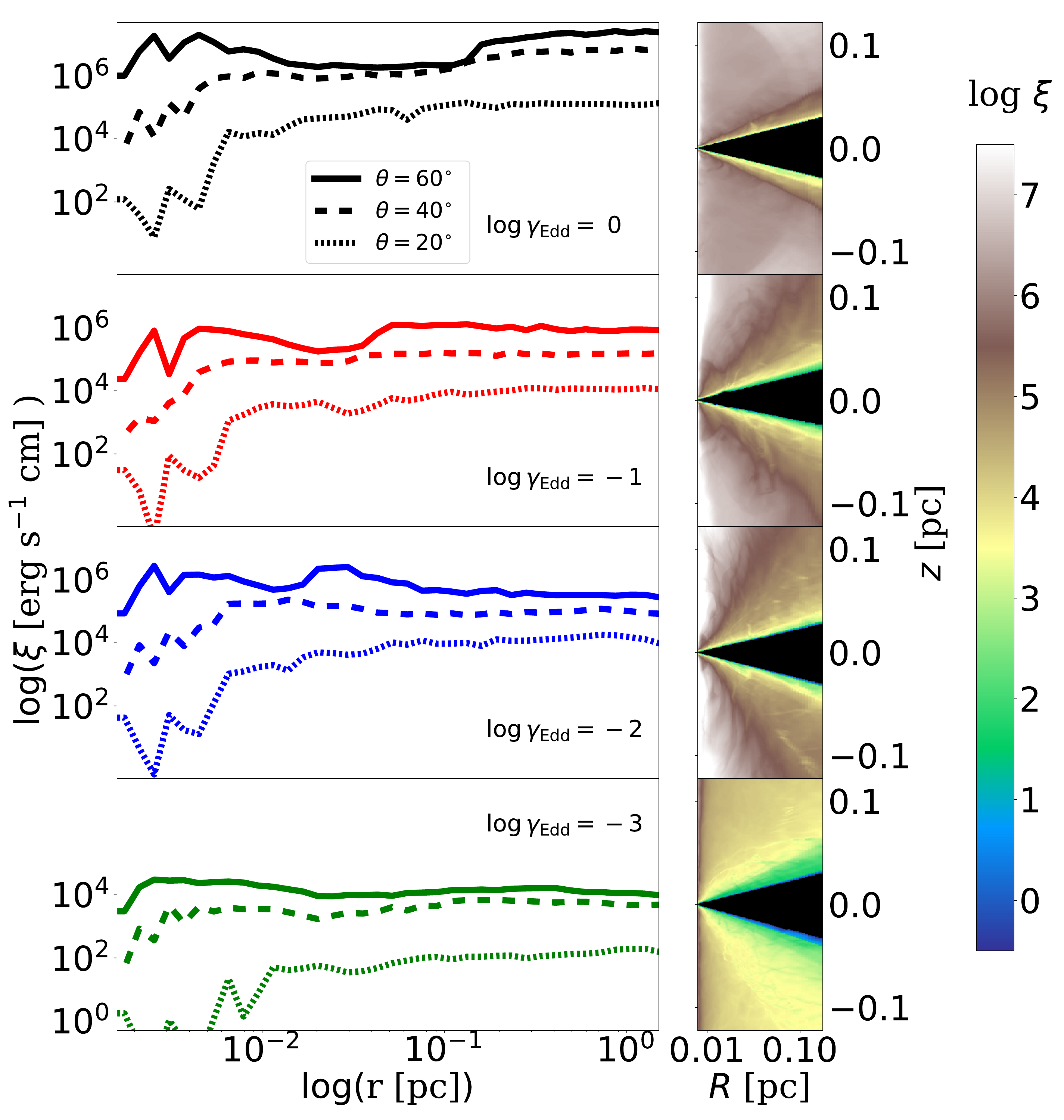}
%\end{interactive}
\caption{
The time-averaged ionization degrees are arranged from top to bottom in descending order of Eddington ratio. 
The left panel plots describe the line-of-sight ionization degree for each angle $\theta=60^{\circ}, 40^{\circ}, and 20^{\circ}$. 
The right plots describe the spatial distribution of the time-averaged mean ionization degree. 
For each Eddington ratio, the black regions in the $R$--$z$ plane represent areas where the column density exceeds $10^{26}$ cm$^{-2}$, i.e. $\xi \sim 0 ~\text{erg s}^{-1}\text{ cm}$.
\label{fig:10}
}
\end{figure}
%%%%%%%%%%%%%%%%%%%%%%%%

Figure \ref{fig:10} shows the line-of-sight and spatial distribution of the density-weighted time-averaged ionization parameter. 
This definition evaluates the ionization degree considering the trace of the shocked shells and the attenuation of the outflowing gas as $\int L_{\rm ion} e^{-\tau(t)} dt /  (r^2 \int n(t) dt)$.
When fixing the angle to $\theta=60^{\circ}$ (solid lines), the ionization parameter decreases as the Eddington ratio decreases. 
Moreover, the ionization parameter decreases at angles closer to the disk.
Therefore, we can estimate the ionization parameter experienced by the averaged shell of the column density and the assumed SED for each Eddington ratio.
\begin{equation}
\xi \sim 10^7 \quad \text{erg s}^{-1}\text{ cm}  \left( \frac{L_{\text{ion}}}{10^{44} \text{erg s}^{-1} } \right) \left( \frac{n}{10^5 \text{ cm}^{-3}}\right)^{-1} \left( \frac{r}{0.01 \text{pc}}  \right)^{-2} \left( \frac{\exp (- m_H N_H \kappa_T)}{1} \right)
\label{eq:xi}
\end{equation}
In the radial profile of Figure \ref{fig:10}, $\xi$ becomes approximately constant for $r \gtrsim 0.1$ pc. 
This is consistent with the power law of number density $n\propto r^{-2}$ in Figure \ref{fig:4}, resulting in $\xi \propto n^{-1} r^{-2} \simeq \text{const}$.

%分布 
We predict the spatial scales corresponding to the observed X-ray absorption and emission lines based on our outflow model.  
For each Eddington ratio, the ionization parameter $\xi(\theta=20^{\circ})$ (dashed lines) is two orders of magnitude smaller than that at $\theta=60^{\circ}$ (solid lines). 
This difference can be attributed to the column density shown in Figure \ref{fig:8}, which is independent of the Eddington ratio and indicates that at $\theta=20^{\circ}$, $N_H \sim 10^{25}$ cm$^{-2}$, i.e. $e^{-\tau} \sim 0.02$.
Emission lines are likely to arise from the re-emission of gas at small angles (\citealt{2023MNRAS.526.2717W}) and can be associated with the BLR (\citealt{2011A&A...525L...8C,2017ApJ...846..154C,2021ApJ...920...30N,2024Univ...10...29N}). 
However, X-ray absorption lines require angles where $N_H \lesssim 10^{24}$ cm$^{-2}$. 
According to Figure \ref{fig:10}, our model predicts that gas of the escape velocity with $\xi<10^4 ~\text{erg s}^{-1}\text{cm}$ within $0.01$ pc contributes to the absorption lines as some WAs. 
As shown in Figure \ref{fig:5}a, sub-parsec scale outflows are distributed across a wide range of ionization parameters, with velocities ranging from $10^3$ to $10^4$ km s$^{-1}$.
\cite{2003ApJ...597..832K,2007ApJ...663..799H,2014MNRAS.441.2613L} reported that the WAs in the ionization range of $2 \lesssim \log \xi \lesssim 4$ are thermally unstable gas on the thermal equilibrium \citep[e.g.][]{2018ApJ...868..111S,2024ApJ...960..120S}.
The shocked-shells outflow may be related to the density and temperature in this ionization state.

Our results suggest that the WAs probes only a part of the dusty outflow. 
The outflowing gas at parsec scale in Figure \ref{fig:10} exhibits a high ionization parameter of $\xi \gtrsim 10^4$ erg s$^{-1}$ cm with $v_{\rm outflow} \gtrsim 10^4$ km s$^{-1}$.  
Not only does the high $\xi$ make the WAs detection difficult \citep{2001ApJS..133..221K}, but the low gas density at parsec scale further complicates this \citep{2007ApJ...663..799H,2014MNRAS.441.2613L}. 
However, with improved detector sensitivity, the WAs with $\xi > 10^4$ ($T > 10^{6.5}$ K) may become detectable.

WAs predicted by the dusty outflows differ from UFOs in terms of velocity, ionization degree, and column density. 
\cite{2014MNRAS.441.2613L} demonstrated that the dependence of column density on ionization degree differs between UFOs and WAs \citep[see also,][]{2024ApJS..274....8Y}. 
They emphasized the significance of the UFOs as outflows originating from the accretion disk scale. 
We discussed that the observational discrepancies observed in the obscuration fraction in \S \ref{sec:sub53} suggest the necessity of outflows within the sub-parsec scale. 
These issues will likely be elucidated in future studies that consider the interaction between dusty outflows and UFOs originating from the accretion disk scale.

\subsection{Wind classification}\label{sec:sub55}

Observations of the absorption line profiles with the high-resolution spectrometers on XRISM and Athena would directly constrain the driving processes \citep[see,][]{2023arXiv230210930G}. 
This is evident from the report of multiple components in the Fe K$_\alpha$ emission line using XRISM \citep{2024arXiv240814300X}.
We classified the radiation-driven and thermal winds in Figure \ref{fig:3}.
Magnetic fields, which we did not consider in our simulations, have been noted as important for sub-parsec accretion processes \citep{2020ApJ...904....9K,2023Sci...382..554I} and sub-parsec outflow from accretion disk scales (\citealt{2005ApJ...631..689E,2010ApJ...715..636F,2024ApJ...968...70F}, and also \citealt{2019Galax...7...13K}. 
\cite{2022ApJ...940....6F} suggested that comparing characteristic absorption profiles using UFOs can elucidate the driving mechanism. 
These profiles create an asymmetric shape with an extended blue wing for MHD-driven wind, while radiation driving tends to produce a red wing profile due to the terminal velocity. 
Because of slow motion, the thermal-driven wind allows for a relatively narrow line with little asymmetry. 
Additionally, the warm absorber at low $v_{\rm outflow}$ may be useful for differentiating between profiles, with studies relating to the ionization parameter of magnetic fields \citep{2019ApJ...884..111K}, radiation-driven winds \citep{2022ApJ...925...55O}, and thermally-driven winds \citep{2019MNRAS.489.1152M,2021ApJ...914..114G}.

The difference in wind-driving mechanisms is potentially linked to the spatial distribution of the outflowing gas.
While dusty outflows occur the intermittent shell-like structures, several simulation studies have demonstrated the formation of clumpy outflows. 
\cite{2013PASJ...65...88T,2018PASJ...70...22K}, using radiation hydrodynamics simulations, revealed that under high Eddington ratios, clumpy structure form due to radiation hydrodynamic instabilities \citep{2001ApJ...549.1093S, 2014PASJ...66...48T}. 
The clumpy gases have an optical depth of around 1, as determined by gas opacity of the electron scattering.
\citet{2017MNRAS.467.4161D,2020ApJ...893L..34D,2022ApJ...931..134W} investigated the clumpiness due to thermally unstable photoionized gas related as the WAs \citep{2021ApJ...914..114G}. 
\cite{2023MNRAS.525.2668S} showed dusty clumps driven by resonant drag instability \citep{2018MNRAS.480.2813H}, where dust and gas decouple motion. 
Note that they discussed different gas densities and spatial scales.

The density structures in the long-term variability have been discussed by WAs \citep{2016MNRAS.457..510S,2018ApJ...868..111S} and obscuration \citep{2002ApJ...571..234R,2014MNRAS.437.1776M,2014MNRAS.439.1403M,2023NatAs...7.1282R}.
In our model, it is expected that shell formation driven by dusty outflows will manifest observable signatures in the time-variable column density. 
Based on the velocity of the dust-free region (at $10^{-2}$ pc) shown in Figure \ref{fig:4}(a), we predict ten-year-scale variability, i.e. $t_{\rm var}\sim 10 \text{ yr } (v_{\rm outflow}/10^3 \text{ km s}^{-1} )^{-1} (r/ 10^{-2} \text{ pc })$, for X-ray. 
Similarly, the dusty gas at 1 pc varies with a timescale of approximately 100 years for IR.

\section{Summary}  \label{sec:summary}

We investigated the effect of the Eddington ratio on the sub-parsec-scale outflow for the SMBH mass of $10^7$ M$_{\odot}$ via 2D radiation hydrodynamics simulations spatially resolved with the dust sublimation radii from $10^{-3}$ to $1$ parsec. 
Our findings on the sub-parsec-scale outflow are summarized as follows.

%(1) How does the radiation feedback to the sub-parsec-scale gas depend on the Eddington ratio?
(1)
When the Eddington ratio of $\log \gamma_{\rm Edd} > 10^{-3}$, the radiation force overcame the gas pressure. 
This resulted in the stronger outflow and larger dust sublimation radius.
Given $\theta$, the temperature and number density of outflows was almost independent of the Eddington ratio.
However, the density within 0.1 parsec and outflow velocity outside $r_{\rm sub}$ were dependent on the angle because of the gas acceleration of the gas pressure or the dust-radiation force.

%(2) What dynamics govern the radial distribution of the dynamical outflow?
(2) 
We found that the analytical solution could accurately reconstruct the radial profile of the time-averaged outflow velocity in our simulations. 
One of key physical quantities is the critical radius ($r_{\rm cr,g}$) indicating dynamical equilibrium. 
The outflow velocity followed a self-similar solution similar to the canonical stellar wind from supersonic to supersonic. 
However, this radius disappeared within the range of $-3 < \log \gamma_{\rm Edd} < -1$ because of $r_{\rm sub} < r_{\rm cr,g}$.
The other key factor was the dust sublimation radius, which distinguished between dusty and dust-free gases.

%(3) What is the Eddington ratio at which radiation feedback activates on the sub-parsec scale?
(3)
The radiation-driven wind was activated under conditions where $r > r_{\rm sub}$ and $\Gamma_{\rm Edd, d} > 1$. 
In addition, the terminal velocity resulting from the dust-radiation force was proportional to $\gamma_{\rm Edd}^{1/4}$ and $M_{\rm SMBH}^{1/4}$. 
These dynamical structures that are dependent on the Eddington ratio are summarized in Figure \ref{fig:7}.

%(4) How does the dynamical dusty outflow explain the CFs of IR and X-ray observations?
(4)
The dust-free gas exhibited a large contribution to the column density from any angle viewing into the center. 
Eddington ratio dependence was not observed in the dust-free gas and weakly in the dusty gas.
The analytical solution showed that the scale and density of the dust sublimation radius facilitated explain this behavior.  
By converting the $\theta$ dependence to $f_{\rm obs}$, our results for $N_{\rm H}=10^{22}$ cm$^{-2}$ explained the high X-ray obscuration of $f_{\rm obs, X} \sim 0.9$ where $\log \gamma_{\text{Edd}} \lesssim-1.5$ and IR obscuration of $f_{\rm obs, IR} \sim0.5$.
The dusty and dust-free gases obscuring the nucleus were essentially determined by the dust sublimation radius.

These results at the sublimation scale provide important information on the parsec-scale torus and Schwarzschild-scale accretion disk. 
The blowout gas at the dust sublimation scale has the potential to form a dynamical dusty torus, linking to the radiation-driven fountain model described by \citet{2015ApJ...812...82W}. 
In our findings, the gas distribution on the dust sublimation scale contributes to the column density and obscuring fraction describing the Compton thickness. 
By considering contributions from more powerful outflows originating from the Schwarzschild scale, which were not considered in our simulations, a more natural explanation can be presented for the observed $f_{\rm obs}$. 
Therefore, we emphasize the importance of various AGN outflows ranging from the Schwarzschild to the torus scales in the system of galactic nuclei.

\acknowledgments
We thank the anonymous referee for valuable comments and suggestions.
We are very grateful to Ichikawa Kohei for useful comments on the analytics of observed data.
Numerical computations were carried out on Cray XC50 at the Center for Computational Astrophysics, National Astronomical Observatory of Japan. 
For the parameter survey of the numerical model, this study used the computational resources of the supercomputer Fugaku provided by RIKEN through the HPCI System Research Project (Project ID: hp210147, hp210219). 
This study was supported by JSPS KAKENHI grant No. 24K17080 (Y.K.), 19K03918 (N.K.), 20K14525 (M.N.), and 21H04496 (K.W.). 
Y.K. and K.W. were supported by NAOJ ALMA Scientific Research grant No. 2020-14A.

%We thank all the people that have made this AASTeX what it is today.  This
%includes but not limited to Bob Hanisch, Chris Biemesderfer, Lee Brotzman,
%Pierre Landau, Arthur Ogawa, Maxim Markevitch, Alexey Vikhlinin and Amy
%Hendrickson. Also special thanks to David Hogg and Daniel Foreman-Mackey
%for the new "modern" style design. Considerable help was provided via bug
%reports and hacks from numerous people including Patricio Cubillos, Alex
%Drlica-Wagner, Sean Lake, Michele Bannister, Peter Williams, and Jonathan
%Gagne.


\vspace{5mm}
%\facilities{HST(STIS), Swift(XRT and UVOT), AAVSO, CTIO:1.3m,


\software{CANS+ \citep{2019PASJ...71...83M}
          }



%\appendix
\bibliography{bibtexEddCF}
%\begin{thebibliography}{}
%\bibitem[Kato et al.(2023)]{2023ApJ...950...72K} 
%Kato, Y., et al. 2023, \apj, 950, 72. Paper~I.
%\end{thebibliography}

%\bibitem[Kudoh et al.(2023)]{2023ApJ...950...72K} Kudoh, Y., Wada, K., Kawakatu, N., et al.\ 2023, \apj, 950, 72. doi:10.3847/1538-4357/accc2b
\bibliographystyle{aasjournal}



\end{document}